\UseRawInputEncoding
\documentclass[letter,prapplied, amsmath,amssymb,
 reprint,
 author-year,%
superscriptaddress,
longbibliography
]{revtex4-2}

\usepackage{natbib}
\usepackage{cancel}
\usepackage[export]{adjustbox}

\usepackage{enumitem}
\usepackage{float}
\usepackage{graphicx}
\usepackage{amsmath,amsthm,amssymb,amsfonts}
\usepackage{dcolumn}
\usepackage{bm}
\usepackage{gensymb}
\usepackage[normalem]{ulem}
\usepackage{hyperref}
\usepackage[all]{hypcap} 
\usepackage[caption=false]{subfig}
\graphicspath{ {images/} }
\usepackage[colorinlistoftodos]{todonotes}
\usepackage{wasysym}
\usepackage{array}
\usepackage{xcolor}
\usepackage[utf8]{inputenc}
\usepackage{float}
\usepackage{siunitx}
\usepackage{braket}
\usepackage{derivative}
\usepackage[normalem]{ulem}

\newcolumntype{C}{>{\centering\arraybackslash}X} 

\usepackage{cleveref}

\crefname{equation}{Eq.}{Eqs.}
\Crefname{equation}{Equation}{Equations}
\crefname{figure}{Fig.}{Figs.}
\Crefname{figure}{Figure}{Figures}
\crefname{section}{Sect.}{Sects.}
\Crefname{section}{Section}{Sections}
\crefname{table}{Table}{Tables}
\crefname{appsec}{Appendix}{Appendices}
\graphicspath{{Figures/}}

\usepackage{color, colortbl}
\definecolor{Gray}{gray}{1}
\definecolor{orange}{rgb}{1,0.97,0.9}
\definecolor{cyan}{rgb}{0.92,1,1}
\definecolor{red}{rgb}{1,0.9,0.9}
\definecolor{purple}{rgb}{0.5, 0.0, 0.5}

\usepackage{soul}

\usepackage{hyperref}
\usepackage[all]{hypcap}

\usepackage[caption=false]{subfig}

\begin{document}

\title{Universal Jaynes--Cummings Control of an Oscillator}

\author{Jordan Huang}
\email{jah499@scarletmail.rutgers.edu}
\affiliation{Department of Physics and Astronomy, Rutgers University, Piscataway, NJ 08854, USA}
\author{Ethan Kasaba}
\affiliation{Department of Physics and Astronomy, Rutgers University, Piscataway, NJ 08854, USA}
\author{Thomas J. DiNapoli}
\affiliation{Department of Physics and Astronomy, Rutgers University, Piscataway, NJ 08854, USA}
\author{Tanay Roy}
\affiliation{Superconducting Quantum Materials and Systems Center, Fermi National Accelerator Laboratory, Batavia, IL 60510, USA }
\author{Srivatsan Chakram}
\email{schakram@physics.rutgers.edu}
\affiliation{Department of Physics and Astronomy, Rutgers University, Piscataway, NJ 08854, USA}
\date{\today}

\begin{abstract} 

The Jaynes--Cummings (JC) interaction---the coherent exchange of excitations between a two-level system and a harmonic oscillator---is one of the fundamental interactions of quantum optics, realized across platforms such as cavity quantum electrodynamics, trapped ions, mechanical resonators, and superconducting circuits. Although JC interactions and qubit rotations form a universal gate set for oscillator control, practical implementations have not been demonstrated. Here we develop and experimentally demonstrate universal JC-based oscillator control by compiling arbitrary unitary gates into sequences of JC interactions and qubit rotations. In our experiment, the oscillator is realized using a mode of a high quality factor microwave cavity and the ancilla qubit using a superconducting transmon circuit, with the JC interaction implemented by a sideband interaction enabled by the Josephson nonlinearity. The native gates are constructed to be closed below a chosen cutoff photon number, encoding a qudit with suppressed leakage errors, while ancilla relaxation errors are detectable. We further find that the dispersive shift serves as a compilation resource that reduces circuit depths. We demonstrate universal qudit control and implement a single-qutrit gate set with a mean post-selected process fidelity of $96.3\%$, as well as ququart and ququint shift gates. These results establish Jaynes--Cummings control as a practical route to universal oscillator control, enabling programmable bosonic processors across a variety of quantum platforms.

\end{abstract}

\pacs{Valid PACS appear here}
\maketitle

The Jaynes--Cummings (JC) interaction is a cornerstone of quantum optics, forming the basis of landmark demonstrations in cavity quantum electrodynamics (QED) that revealed the quantum nature of light~\cite{raimond2001manipulating}. It provides a foundational paradigm for quantized light-matter and spin-boson interactions, enabling control of bosonic degrees of freedom through coherent coupling to an ancillary two-level system. It is native to a wide 
variety of platforms, including trapped ions~\cite{leibfried2003quantum, fluhmann2019encoding}, circuit 
QED~\cite{wallraff2004strong, reagor2016quantum}, cavity 
magnonics~\cite{tabuchi2015coherent}, and quantum 
acoustodynamics~\cite{chu2017quantum, satzinger2018quantum}. A protocol for arbitrary state preparation of a bosonic mode using the JC interaction was first developed in the seminal work of Law and Eberly~\cite{law1996arbitrary} and subsequently demonstrated experimentally using trapped ions~\cite{ben2003experimental} and 
superconducting circuits~\cite{hofheinz2009synthesizing}.

This framework has since been extended to universal control~\cite{mischuck2013qudit}, including a technique that uses JC rotations that are closed below a chosen photon number~\cite{liu2021constructing},  naturally encoding a qudit in the oscillator. Despite these theoretical advances, universal JC-based control has not been realized experimentally: existing demonstrations have focused on state preparation and restricted classes of gate operations~\cite{valadares2026flux, huang2025fast}.

Superconducting circuits coupled to high-$Q$ microwave cavities provide a powerful setting for oscillator control, combining long coherence times~\cite{milul2023superconducting,kim2025ultracoherent,Oriani2025Nb} with the large coupling strengths and high cooperativities of circuit QED~\cite{blais2021circuit}. This enables access to the strong-dispersive regime of cavity QED, where the addition of a single photon can be spectrally resolved~\cite{schuster2007resolving}. The dispersive interaction enables selective photon number-dependent phase gates which, combined with oscillator displacements, realize universal 
control~\cite{krastanov2015universal, heeres2015cavity,1508.05882, reinhold2020error,fosel2020efficient,landgraf2024fast, wang2016schrodinger, liu2026hybrid}. Achieving state-of-the-art performance typically~\cite{you2025floquet} requires large dispersive shifts, which introduce ancilla-mediated errors that are frequently the dominant error channel. This limits the performance of bosonic quantum error correction~\cite{ofek2016extending, ni2023beating} and becomes increasingly problematic as oscillator coherence times improve~\cite{Oriani2025Nb,milul2023superconducting,kim2025ultracoherent}.

This limitation has motivated approaches that work in the weak-dispersive regime such as those based on ancilla conditional displacements~\cite{eickbusch2021fast,diringer2024conditional, campagne2020quantum,sivak2023real, you2024crosstalk}. However, these require large oscillator displacements that populate states beyond the computational space during the gate, making it challenging to account for higher-order Hamiltonian terms and reach decoherence-limited fidelities. This motivates a return to the JC interaction as the native bosonic control primitive, where leakage suppression is enforced by construction and the gate set is composed from a discrete family of independently calibrated primitives. The JC interaction can be implemented using a sideband 
drive~\cite{pechal2014microwave, rosenblum2018cnot} and can be 
made strong in the weak-dispersive 
regime~\cite{huang2025fast}, mitigating ancilla-mediated errors.



In this work, we introduce and demonstrate ancilla-error-detectable, universal oscillator control using the JC interaction as the native control primitive. We develop efficient decompositions of oscillator-encoded qudit unitaries using sequences of JC and qubit rotations. We find that the dispersive shift, rather than degrading JC-based control as expected, can be used as a compilation resource to accelerate gates. Treating the JC detuning as a control parameter enhances mixing across the oscillator manifold and improves circuit depth scaling with qudit dimension. We use this protocol to experimentally implement a universal single-qudit gate set, demonstrating both Clifford and non-Clifford gates for a qutrit along with ququart and ququint shift gates. We characterize their performance using joint qubit–oscillator Wigner tomography and qudit process tomography.

We implement the JC control protocol in a multimode circuit-QED device, where the oscillator is realized as a mode in a superconducting microwave cavity with millisecond oscillator coherences that is weakly coupled to a transmon~\cite{huang2025fast}, in which the ancilla qubit is encoded in its ground $\ket{g}$ and second-excited $\ket{f}$ states. We charge-drive the transmon and use the Josephson nonlinearity to activate a tunable JC interaction~\cite{pechal2014microwave} given by  
\begin{equation}
    \mathcal{H}=g_\mathrm{JC}\left(e^{i\phi}\sigma_-a^\dag + e^{-i\phi}\sigma_+a\right),
\end{equation}
where $\sigma_-=\ket{g}\bra{f}$, $\sigma_+=\ket{f}\bra{g}$, and $a$ is the lowering operator of the oscillator. By using a sideband drive to implement the JC interaction, we can achieve interaction rates more than an order of magnitude larger than the dispersive shift, with the characteristic $\sqrt{n}$ bosonic enhancement with photon number~\cite{huang2025fast}. 

\begin{figure}
    \centering
    \includegraphics[width=0.95\columnwidth]{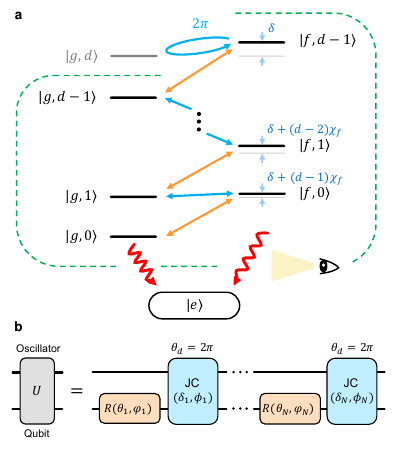}
    \caption{\textbf{Jaynes--Cummings control.} \textbf{(a)} Energy level diagram showing JC (cyan) and ancilla qubit (orange) interactions in a rotating frame where the qubit and oscillator have a JC detuning. The ancilla qubit is encoded in the transmon's $\ket{g}$ and $\ket{f}$ states. The JC interaction is implemented using a sideband drive detuned by $\delta$ from the $\ket{f,d-1} \leftrightarrow \ket{g,d}$ transition, where $d-1$ is a chosen photon-number cutoff. Corresponding transitions for other photon numbers are further detuned by dispersive shifts. The pulse duration is chosen to implement a $2\pi$ rotation on the cutoff transition, closing the dynamics to the subspace within the green box. Ancilla qubit relaxation errors send the state to $\ket{e}$, which is outside of the computational subspace, allowing for their detection. \textbf{(b)} Circuit representation of the JC protocol. An arbitrary unitary can be approximated by a sequence of JC and ancilla qubit rotations.}
    \label{fig1}
\end{figure}

The JC protocol decomposes arbitrary unitary gates into alternating sequences of ancilla qubit rotations and JC rotations. The action of each operation is illustrated in the level diagram in Fig.~\ref{fig1}(a). Photon-number-unselective ancilla qubit rotations $R(\theta,\varphi)=\exp{(i\frac{\theta}{2}(\sigma_x\sin\varphi+\sigma_y \cos\varphi))}$ between $\ket{g}$ and $\ket{f}$ are implemented using a composite sequence composed of transmon $\ket{g}$-$\ket{e}$ rotations in between two $\ket{e}$-$\ket{f}$ $\pi$ rotations; details are given in Sec.~S1 of the 
Supplementary Information (SI). JC rotations couple doublets $\{\ket{f,n-1},\ket{g,n}\}$ with 
photon-number-dependent rates. In our protocol, these rotations are 
engineered to execute a $2\pi$ rotation on a chosen cutoff transition $\ket{f,d-1}\leftrightarrow\ket{g,d}$, closing the dynamics to the computational subspace spanned by $\{\ket{0},\ldots,\ket{d-1}\}$ and 
naturally encoding a $d$-level qudit in the oscillator. This closed subspace is indicated by the dashed green border in Fig.~\ref{fig1}(a). The $\sqrt{n}$ scaling 
of the JC interaction rate produces nontrivial mixing in the lower doublets $\{\ket{f,n-1},\ket{g,n}\}$ for $n < d$. Combining these primitives gives the circuit decomposition illustrated in Fig.~\ref{fig1}(b), which compiles arbitrary unitaries into alternating ancilla-qubit and JC rotations. While this construction is universal~\cite{liu2021constructing}, efficient decompositions for arbitrary unitaries have not been demonstrated, which we address here. 

Our control protocol has several advantages. The closed 
dynamics strongly suppresses leakage errors and eliminates enhanced photon loss from populating higher Fock states. By confining the state dynamics to the computational subspace, it is also numerically efficient to simulate and model errors. Encoding the ancilla qubit in the ground and second-excited states of the transmon enables the detection of ancilla relaxation errors: a decay $\ket{f}\rightarrow\ket{e}$ or excitation $\ket{g}\rightarrow\ket{e}$ shelves the state in $\ket{e}$, where it remains unaffected by subsequent gate layers: $\ket{e}$ returns to itself after each ancilla rotation and is dark to the JC interaction. A single ancilla relaxation error can therefore be detected at the end of any sequence. See Sec.~S8 of the SI for a comparison to other control protocols. 

A static JC interaction leads to photon-number-dependent shifts of the qubit frequency in the dispersive regime, which was not considered in Ref.~\cite{liu2021constructing}. These shifts are significant in circuit-QED platforms and described, in our case, by the Hamiltonian $\mathcal{H}=\chi_f a^\dag a\sigma_z/2$. Here, $\sigma_z=\ket{f}\bra{f} - \ket{g}\bra{g}$ and $\chi_f$ is the dispersive shift of the transmon $\ket{f}$ state. While the $\sqrt{n}$ scaling of the JC interaction rate guarantees nontrivial mixing across Fock states in the idealized $\chi_f = 0$ limit, the dispersive shift in realistic circuit-QED devices has a significant impact on the dynamics, set by the relative strength of the JC coupling $g_{\mathrm{JC}}$ to the dispersive shift $\chi_f$. For high-dimensional qudits, the construction of Ref.~\cite{liu2021constructing} degrades: the lower doublets $\{\ket{f,n-1},\ket{g,n}\}$ for $n<d$ are detuned from the cutoff transition $\ket{f,d-1}\leftrightarrow\ket{g,d}$ by $(n-(d-1))\chi_f$, suppressing the mixing necessary for efficient unitary synthesis. We address this challenge by introducing a controlled detuning $\delta$ of the JC interaction from the cutoff transition, as shown in Fig.~\ref{fig1}(a). We still enforce a $2\pi$ rotation on the cutoff transition and find that the detuning not only remedies the loss of mixing at finite $\chi_f$, but also reduces the required circuit depth below that achievable in the $\chi_f = 0$ limit, leading to faster gates at fixed dispersive shift.

\begin{figure}
    \centering
    \includegraphics[width=\columnwidth]{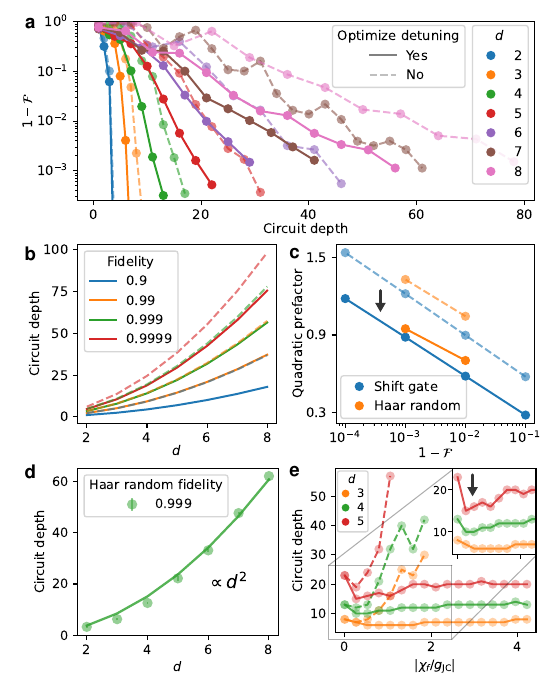}
    \caption{\textbf{Scaling analysis.} \textbf{(a)} Gate infidelity for a shift gate as a function of circuit depth, for different qudit dimensions. The solid and dotted lines indicate optimization with and without JC detuning, respectively. \textbf{(b)} Circuit depth required to reach a target fidelity as a function of qudit dimension, fit to a quadratic function. \textbf{(c)} Quadratic prefactor of the shift gate (blue) and Haar random gate (orange) circuit depths as a function of target infidelity, showing logarithmic scaling and a reduction with detuning optimization. \textbf{(d)} For Haar random unitaries, the circuit depth required to reach a target fidelity as a function of qudit dimension displays a quadratic scaling. \textbf{(e)} Circuit depth required to reach $> 99\%$ fidelity as a function of $|\chi_f / g_{\mathrm{JC}}|$, for different qudit dimensions. Optimizing with a JC detuning leads to a significant improvement in circuit depth.} 
    \label{fig2}
\end{figure}

To compile target qudit unitaries with our JC protocol, we use multistart gradient descent with automatic differentiation to optimize the circuit parameters $\{\theta_i, \varphi_i, \delta_i,\phi_i\}$, where $i$ is the circuit layer, $\theta$ is the qubit's angle of rotation, $\delta$ is the JC interaction's detuning, and $\varphi$ and $\phi$ are the azimuthal angles of the rotation axes for the qubit and JC rotation, respectively. More details can be found in Sec.~S4 of the SI.

As a benchmark, we focus on the qudit analog of the Pauli-$X$ operator, referred to as a shift gate, where $X=\sum_{j=0}^{d-1}\ket{j+1\;(\mathrm{mod}~d)}\bra{j}$. In Fig.~\ref{fig2}(a), we study its optimizer infidelity as a function of the number of circuit layers used for optimization, for various qudit dimensions $d$. The dashed curves correspond to the protocol with fixed detuning $\delta = 0$, while the solid curves show the case in which $\delta$ is treated as an optimization parameter. Numerically, we find that the circuit depth required to reach a given target infidelity scales as $\mathcal{O}(d^2)$ (shown in Fig.~\ref{fig2}(b)), which is consistent from a parameter counting argument for a $d$-dimensional unitary. We fit these results to extract the quadratic prefactor of the scaling, shown in Fig.~\ref{fig2}(c), as a function of the target infidelity. This prefactor depends logarithmically on the infidelity. With our experimental parameters of $|\chi_f/g_{\mathrm{JC}}| \approx 0.5$, the prefactor is reduced by a factor of $1.4$ for a $99\%$ fidelity shift gate when the detuning is allowed to vary.

\begin{figure*}
    \centering
    \includegraphics[width=0.95\linewidth]{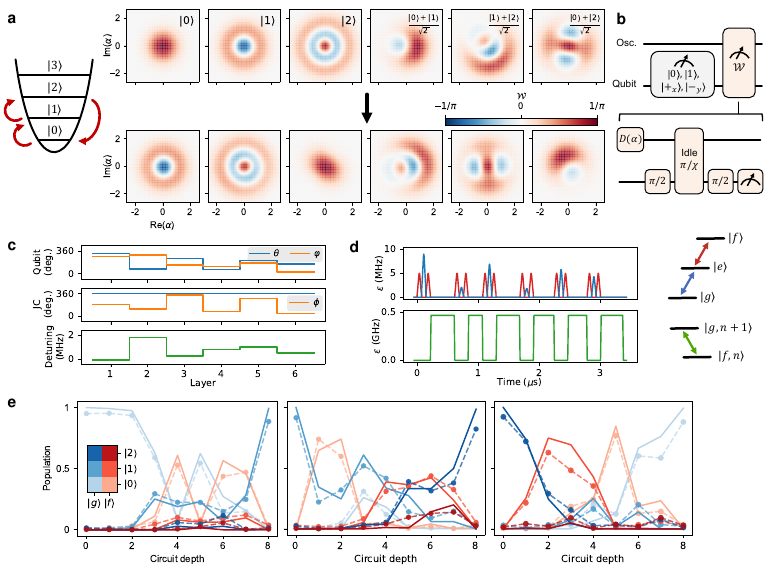}
   \caption{\textbf{Qutrit shift gate.} \textbf{(a)} Illustration of a qutrit shift gate in an oscillator. Experimentally reconstructed Wigner functions are shown for six input states (top row) and the corresponding output states after the shift gate (bottom row). \textbf{(b)} Sequence used for joint qubit-oscillator tomography. The oscillator state is measured using Wigner tomography after postselecting on four qubit basis states. \textbf{(c)} Angles of the ancilla qubit and JC rotations, and JC detunings, for a qutrit shift gate. \textbf{(d)} Pulse sequence corresponding to the rotation angles in \textbf{(c)}. \textbf{(e)} Populations $P(\ket{g,n})$ and $P(\ket{f,n})$ as a function of circuit layer, measured using joint qubit--oscillator tomography for a qutrit shift gate initialized in $\ket{g0}$, $\ket{g1}$, and $\ket{g2}$. The markers show the experimentally measured populations and the lines display the simulated decoherence-free evolution.}
    \label{fig3}
\end{figure*}

To examine the effect of the dispersive shift, we perform circuit optimization for different values of $\chi_f/g_{\mathrm{JC}}$. For each choice of $\chi_f/g_{\mathrm{JC}}$, we compute the fidelity as a function of the number of layers and extract the minimum depth required to reach a given target fidelity (shown in Fig.~\ref{fig2}(e)). In the $\delta = 0$ case, the number of layers needed for a given fidelity increases with increasing $\chi_f/g_{\mathrm{JC}}$, reflecting reduced mixing among the lower photon number states at large dispersive shifts. By contrast, allowing $\delta$ to vary as an optimization parameter stabilizes the required depth. Furthermore, there is a $\chi_f/g_{\mathrm{JC}}$ regime where the required depth outperforms the case where $\chi_f=0$ when the detuning is allowed to vary, indicated by the arrow in the inset in Fig.~\ref{fig2}(e). This demonstrates that the dispersive shift, rather than merely degrading performance, can serve as a compilation 
resource in experimentally relevant circuit-QED parameter regimes.

The $\mathcal{O}(d^2)$ circuit-depth scaling shown for shift gates is robust for other qudit gates, as demonstrated by the similar depth scaling required to implement Haar-random unitaries, shown in Fig.~\ref{fig2}(d). The prefactor of the quadratic dependence is comparable to that of the shift gate (Fig.~\ref{fig2}(c)). Furthermore, the bosonic enhancement of the JC interaction rate with photon number means that the gate time scales as $\mathcal{O}(d^{3/2})$, since JC rotations at the cutoff transition for a $d$-dimensional qudit are $\sqrt{d}$ times faster. 

Since the unitary operations presented in this work rely on interference between JC and ancilla qubit rotations, we must accurately model phase corrections that arise from the physical implementation of the gate set, which we implement through sideband and resonant drives on a transmon. In particular, we must correct for driven Stark shifts and dispersive interactions that modify the sideband and transmon transition frequencies, which we detail in Sec.~S2 of the SI. 

We now present experimental demonstrations of our JC control protocol, starting with a qutrit shift gate $X$ in the oscillator, illustrated in Fig.~\ref{fig3}(a). A JC interaction is applied with a sideband drive on resonance with the $\ket{f,2} \leftrightarrow \ket{g,3}$ transition, or near 
resonance with a small programmable detuning $\delta$, with its duration chosen to implement a $2\pi$ rotation. We implement circuits with an optimized decoherence-free fidelity of $>99\%$, requiring 6 (8) layers with (without) detuning optimization. 
The ancilla and JC rotation angles and JC detunings for the detuning-optimized 6-layer shift gate circuit, along with the corresponding pulses, are shown in Fig.~\ref{fig3}(c) and (d).

\begin{figure*}
    \centering
    \includegraphics[width=\linewidth]{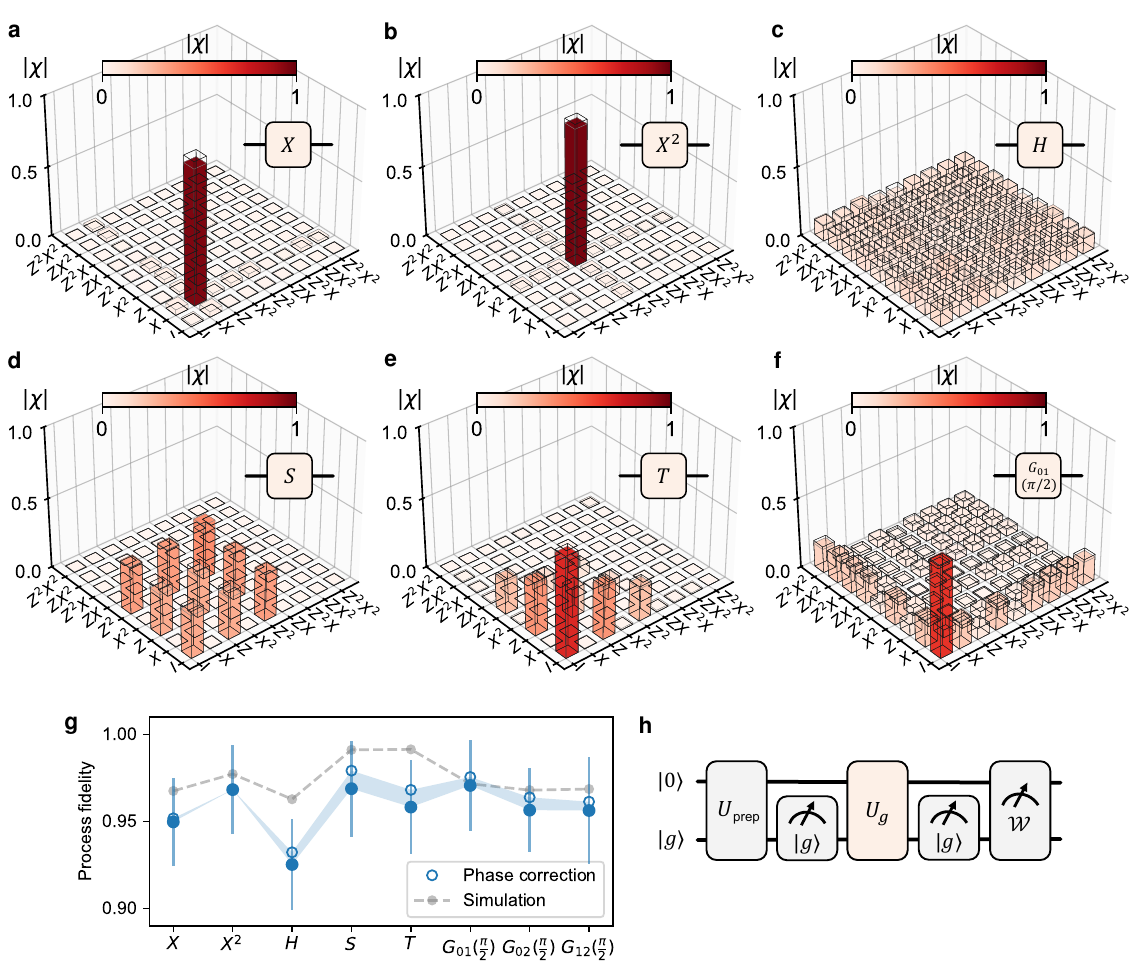}
    \caption{\textbf{Process tomography of universal qutrit gates.} Absolute value of the process matrix for a qutrit \textbf{(a)} shift gate $X$, \textbf{(b)} $X^2$ gate, \textbf{(c)} Hadamard gate, \textbf{(d)} phase gate $S$, \textbf{(e)} $T$ gate, and \textbf{(f)} $\pi/2$ Givens rotation in the $\ket{0}$-$\ket{1}$ subspace. The solid black lines indicate the ideal gate. The full process matrices are shown in Figs.~S9 and~S10 of the SI. \textbf{(g)} Process fidelities of the universal qutrit gates after post-selection are shown as solid blue markers. Grey markers indicate results of Lindblad master equation simulations. The blue hollow markers show the process fidelity obtained when the implemented gate is compared to the target gate followed by a qudit phase gate,  $R(\varphi)U_\mathrm{target}$, where $R(\varphi)=\sum_j e^{ij\varphi}\ket{j}\bra{j}$. Error bars on the fidelities are obtained by extracting errors in the density matrix reconstruction from Wigner tomography following Ref.~\cite{huang2025fast}, and propagating them through the process matrix reconstruction. \textbf{(h)} Sequence used for qutrit process tomography. $U_\mathrm{prep}$ is the gate used to prepare the input state for process tomography and $U_g$ is the gate being characterized.}  
    \label{fig4}
\end{figure*}

We first characterize the shift gate's action on different input states $\ket{g}\otimes\ket{\psi}$ by measuring the input and output states using Wigner tomography, shown in Fig.~\ref{fig3}(a). To obtain a more detailed characterization, we implement a joint qubit-oscillator Wigner tomography 
protocol~\cite{vlastakis2015characterizing}. As illustrated in Fig.~\ref{fig3}(b) and detailed in Sec.~S9 
of the SI, the joint qubit-oscillator density matrix can be reconstructed from Wigner tomography of the oscillator state conditioned on measuring the qubit along different axes. The population evolution of the shift gate as a function of circuit layer is shown in Fig.~\ref{fig3}(e) for an 8-layer circuit without detuning optimization. The markers show the experimentally measured populations, while the lines display the ideal decoherence-free evolution. We perform this characterization for the system initialized in $\ket{g0}$, $\ket{g1}$, and $\ket{g2}$ and find good agreement between 
experiment and theory.

To demonstrate universal qudit control, we implement a representative set of single-qutrit gates (6-layer circuits optimized to $>99\%$) and benchmark them with qudit process tomography~\cite{morvan2021qutrit,Roy2023two_qutrit}. The protocol is shown in Fig.~\ref{fig4}(h) and detailed in Sec.~S10 of the SI. We prepare a tomographically complete set of qutrit input states using the analytic sideband-based state preparation sequences of Ref.~\cite{huang2025fast} and the JC protocol. After applying the gate, we measure the output states conditioned on the transmon being in $\ket{g}$ and use the input and output states to reconstruct the process matrix.

We first implement the shift gate $X$, the squared shift gate $X^2$, and a qutrit Hadamard (or Fourier) gate $H = \frac{1}{\sqrt{3}}\sum_{j,k =0}^2 \omega^{jk}\ket{j}\bra{k}$, where $\omega = e^{2\pi i/3}$. The absolute values of the corresponding experimentally obtained process matrices are shown in Fig.~\ref{fig4}(a)--(c). The $X$ and $H$ gates can generate a $Z$ gate ($Z = HXH^\dagger$), and therefore the demonstrated gates already generate the qutrit Pauli group. To generate the qutrit Clifford group, we implement a quadratic phase gate $S = \sum_{j=0}^{2} \omega^{j(j+1)/2}\ket{j}\bra{j} = \mathrm{diag}(1,\omega, 1)$, whose measured process matrix is shown in Fig.~\ref{fig4}(d). The gate $S$, together with $H$, generates all 216 elements of the qutrit Clifford group. Our protocol can efficiently generate all these qutrit Pauli and Clifford gates directly with similar gate depth, as discussed in Sec.~S6 of the SI.

To move beyond Clifford gates and demonstrate universal qutrit control~\cite{morvan2021qutrit, goss2022high}, we implement four non-Clifford operations. First, we realize the qutrit analog of a $T$ gate, $T = \mathrm{diag}(1, \xi, \xi^8)$ with $\xi = e^{2\pi i/9}$, whose process matrix is shown in Fig.~\ref{fig4}(e). Clifford gates combined with the $T$ gate enable universal qutrit control. Other non-Clifford operations can also be directly realized with the JC protocol, with gate depths comparable to those of the Clifford gates. As a further demonstration of non-Clifford control, we implement $\pi/2$ Givens rotations between all pairs of qutrit states~\cite{brennen2005givens}. The process matrix for a $\pi/2$ Givens rotation between $\ket{0}$ and $\ket{1}$ is shown in Fig.~\ref{fig4}(f), with the rest shown in the SI. The fidelities of the qutrit gates after post-selection are summarized in Fig.~\ref{fig4}(g). We achieve a mean post-selected process fidelity of $95.7\%$ over a representative sample of universal qutrit gates. When allowing for an additional qudit phase gate correction on top of each target gate, the mean fidelity is $96.3\%$, indicated by the blue hollow markers in Fig.~\ref{fig4}(g). This is in close agreement with Lindblad master equation simulations (see Sec.~S11 of the SI for the error budget). The residual phases arise from the post-selection measurement and unaccounted oscillator and transmon Stark shifts and can be corrected in subsequent operations~(see Sec.~S10 and ~S3 of the SI). Further improvements are expected from including these additional Stark shift corrections directly in the optimizer.

To demonstrate our protocol in higher qudit dimensions, we implement ququart ($d=4$) and ququint ($d=5$) shift gates. We characterize these gates using Wigner tomography on a set of Fock-state inputs and outputs. As a fidelity proxy, we compute the state fidelity between the measured output state and the ideal output obtained by applying the target gate to the measured input state, averaged over all inputs. We achieve post-selected fidelities of $95.3\%$ and $94.2\%$ for the ququart and ququint shift gates, respectively. The corresponding analysis is detailed in Sec.~S7 of the SI.

While qudit control has been demonstrated in transmons~\cite{morvan2021qutrit,goss2022high,liu2023performing,champion2025efficient} and trapped ions~\cite{ringbauer2022universal}, these approaches rely on spectrally resolving individual transitions and implementing sequences of $SU(2)$ rotations between selected pairs of levels. A generic $SU(d)$ unitary can be decomposed into $\mathcal{O}(d^2)$ such two-level rotations, but restricting to rotations between adjacent levels increases the required circuit depth to $\mathcal{O}(d^3)$~\cite{liu2023performing}. Alternative constructions achieve $\mathcal{O}(d)$ depth at the cost of applying $\mathcal{O}(d)$ simultaneous drives~\cite{champion2025efficient}, similar to oscillator control with an optimized SNAP-displacement gate set~\cite{fosel2020efficient, landgraf2024fast}. In contrast, our protocol achieves $\mathcal{O}(d^2)$ depth with $\mathcal{O}(d^{3/2})$ gate time scaling for an oscillator-encoded qudit without requiring spectrally resolved rotations or simultaneous multi-drive control. Moreover, because we encode the qudit in an oscillator, photons emitted across different qudit transitions are indistinguishable to the environment. Consequently, the effective error model of our qudit is that of an oscillator, enabling simpler qudit error correction based on bosonic codes~\cite{brock2025quantum, michael2016new}.

In conclusion, we have developed and experimentally demonstrated JC-based universal control of an oscillator. Our gate set is constructed so that the dynamics are closed within the chosen computational subspace, enabling efficient parameter optimization, minimizing leakage errors, and simplifying error modeling. We encode the ancilla qubit in the $\ket{g}$-$\ket{f}$ manifold of a transmon, allowing for the detection of ancilla relaxation errors. Furthermore, we harness the dispersive shift by using the JC detuning as a control parameter and demonstrate a gate compilation speedup. We show that the JC protocol achieves high-fidelity qudit gates with a circuit depth that scales quadratically with the qudit dimension. We implement these protocols in a weakly dispersive multimode cavity--transmon system and realize a universal single-qutrit gate set with a mean post-selected process fidelity of $96.3\%$. We also demonstrate ququart and ququint shift gates with estimated post-selected gate fidelities of around $95\%$, highlighting the advantage of a gate construction that can detect ancilla errors.

In the present device, individual sideband pulses are coherence-limited to fidelities of around $99\%$ by the modest $55~\mu\mathrm{s}$ transmon coherence time. Having state-of-the-art transmon coherence times of $T_1, T_2 \sim 500~\mu s$ ~\cite{bland2025millisecond} will allow our qutrit gates to reach fidelities above $99.3\%$ with post-selection. The JC protocol is also applicable in other circuit-QED architectures, where JC interactions can be implemented by flux-tuning a qubit into resonance~\cite{hofheinz2009synthesizing,
valadares2026flux}, parametric qubit-frequency modulation~\cite{Strand2013,naik2017random}, and using tunable couplers to mediate interactions between qubits and oscillators~\cite{lu2017universal, vool2018driving, li2025cascaded, maiti2025linear}. Our JC protocol also enables generation of arbitrary unitaries within the joint subspace spanned by $\{\ket{g}, \ket{f}\}\otimes\{\ket{0}, \ldots, \ket{d-1}\}$, allowing ancilla-conditioned gates on the oscillator. This provides a clear path for extending our control scheme to a multimode setting~\cite{chakram2020seamless,chakram2020multimode} to realize entangling gates between oscillators. These results establish JC control as a practical and hardware-efficient protocol for universal gates in bosonic processors, providing a foundation for programmable bosonic quantum processors and qudit instruction sets.

\section*{Acknowledgements}

We acknowledge helpful discussions with Liang Jiang, Ming Yuan, Alexander Soudakov, Victor Batista, David Schuster, and Robert Schoelkopf. This work is supported by the Army Research Office under Grant Number W911NF-23-1-0096 and W911NF-23-1-0251, and by the U.S. Department of Energy, Office of Science, National Quantum Information Science Research Centers, Superconducting Quantum Materials and Systems Center (SQMS), under Contract No. 89243024CSC000002.

\bibliography{references}

\end{document}


\title{Supplementary Information: Universal Jaynes--Cummings Control of an Oscillator}

\author{Jordan Huang}
\affiliation{Department of Physics and Astronomy, Rutgers University, Piscataway, NJ 08854, USA}
\author{Ethan Kasaba}
\affiliation{Department of Physics and Astronomy, Rutgers University, Piscataway, NJ 08854, USA}
\author{Thomas J. DiNapoli}
\affiliation{Department of Physics and Astronomy, Rutgers University, Piscataway, NJ 08854, USA}
\author{Tanay Roy}
\affiliation{Superconducting Quantum Materials and Systems Center, Fermi National Accelerator Laboratory, Batavia, IL 60510, USA}
\author{Srivatsan Chakram}
\affiliation{Department of Physics and Astronomy, Rutgers University, Piscataway, NJ 08854, USA}
\date{\today}

\maketitle

\clearpage
\onecolumngrid
\tableofcontents
\clearpage
\twocolumngrid

\section{Qutrit-oscillator Hamiltonian for Jaynes--Cummings control} \label{app:effective_hamiltonian}

Our implementation of Jaynes--Cummings (JC) control is based on a qutrit-oscillator Hamiltonian,
$\mathcal{H} = \mathcal{H}_{0} + \mathcal{H}_{ge} + \mathcal{H}_{ef} + \mathcal{H}_{\mathrm{JC}}$. The derivation of this model from a transmon-oscillator system is detailed in Section \ref{sec:derive_ham}.
Here, the qutrit levels are labeled by $\{\ket{g}, \ket{e}, \ket{f}\}$ and $\mathcal{H}_{0}$ is the undriven dispersive Hamiltonian: 
\begin{equation}
    \mathcal{H}_\text{0} = \chi_e a^\dag a\ket{e}\bra{e} + \chi_f a^\dag a \ket{f} \bra{f},
    \label{undriven_Hamiltonian}
\end{equation}
where $\chi_e$ and $\chi_f$ are the dispersive couplings. The driven interactions, all implemented by microwave drives on the 
transmon, are $\mathcal{H}_{ge}$, which couples $\ket{g}$ and $\ket{e}$,
\begin{equation}
    \mathcal{H}_{ge} = \epsilon_{ge}(t) e^{i(\varphi_1+\varphi_{ge})}\ket{g}\bra{e} + \text{H.c.},
    \label{ge_drive}
\end{equation}
$\mathcal{H}_{ef}$, which couples $\ket{e}$ and $\ket{f}$,
\begin{equation}
    \mathcal{H}_{ef} = \epsilon_{ef}(t) e^{i(\varphi_2+\varphi_{ef})}\ket{e}\bra{f} + \text{H.c.},
    \label{ef_drive}
\end{equation}
and $\mathcal{H}_\mathrm{JC}$, which is a Jaynes--Cummings interaction that couples the states $\ket{f,n}$ and $\ket{g,n+1}$:
\begin{equation}
\begin{split}
    \mathcal{H}_{\mathrm{JC}} &=  \bigl(\epsilon_\text{sb}(t) e^{i(\varphi_3+\varphi_\text{sb}+(\Delta_f-\Delta_\mathrm{osc}+(d-1)\chi_f+\delta)t)}\ket{g}\bra{f}a^\dag \\
    &\phantom{=}\;+ \text{H.c.}\bigl) + \Delta_e(\epsilon_\text{sb})\ket{e}\bra{e} + \Delta_f(\epsilon_\text{sb})\ket{f}\bra{f}\\
    &\phantom{=}\; + \Delta_\mathrm{osc}(\epsilon_{sb}){a}^\dag{a}.
\end{split}
\label{jc_hamiltonian}
\end{equation}  
Here $\varphi_{ge}$, $\varphi_{ef}$, and $\varphi_{\text{sb}}$ are controllable phases associated with each drive. The Jaynes--Cummings interaction is implemented through a sideband drive, where $\delta$ is a controllable drive detuning and $\Delta_e$, $\Delta_f$, and $\Delta_\mathrm{osc}$ are driven Stark shifts of the $\ket{e}$, $\ket{f}$, and oscillator frequencies arising from it (see Sec.~\ref{sec:derive_ham}).
$\varphi_1$, $\varphi_2$, and $\varphi_3$ are static offset phases associated with the drives that can arise from differences in the reference phases of their generators. They can be set to zero by absorbing them into the definition of the basis states by the unitary transformation  $U=\exp\left({-i\varphi_1\ket{g}\bra{g}+i\varphi_2\ket{f}\bra{f}+i(-\varphi_3+\varphi_2)a^\dag a}\right)$. 

With the offset phases set to zero, we now consider the unitary evolution generated by our three driven interactions. The first is $\mathcal{H}_{ge}$, which generates a rotation within the $\ket{g}$-$\ket{e}$ manifold. We consider a square pulse of length $\tau$ with amplitude $\epsilon_{ge}$. Since the Hamiltonian is block diagonal, the $n$-photon block $\{\ket{g,n}, \ket{e,n}, \ket{f,n}\}$ of the unitary is
\begin{align}
\begin{split}
    &R_{ge}^{(n)}(\theta, \varphi_{ge}) = \begin{pmatrix} 
        e^{-i \delta_{ge}\tau} & 0 & 0 \\ 
        0 & e^{-i \delta_{ge}\tau} & 0 \\
        0 & 0 & e^{-in\chi_f \tau}
    \end{pmatrix} \\ 
    &\times\begin{pmatrix}
        c_n + i\frac{\delta_{ge}}{\sqrt{\delta_{ge}^2+\epsilon_{ge}^2}}s_n & -ie^{i\varphi_{ge}}\frac{\epsilon_{ge}}{\sqrt{\delta_{ge}^2+\epsilon_{ge}^2}}s_n & 0 \\
        -ie^{-i\varphi_{ge}}\frac{\epsilon_{ge}}{\sqrt{\delta_{ge}^2+\epsilon_{ge}^2}}s_n & c_n - i\frac{\delta_{ge}}{\sqrt{\delta_{ge}^2+\epsilon_{ge}^2}}s_n & 0 \\
        0 & 0 & 1
    \end{pmatrix},
\end{split}
\label{Q_ge_rotation}
\end{align}
where $\delta_{ge}=n\chi_e/2$, $\theta_n = \theta\sqrt{\delta_{ge}^2+\epsilon_{ge}^2}/\epsilon_{ge}$, and $\theta = 2\epsilon_{ge} \tau$. Here, and in the expressions that follow, $c_n = \cos\frac{\theta_n}{2}$ and $s_n = \sin\frac{\theta_n}{2}$ are used as shorthand.

Similarly, $\mathcal{H}_{ef}$ generates a rotation within the $\ket{e}$-$\ket{f}$ manifold. We consider a square pulse of length $\tau$ with amplitude $\epsilon_{ef}$. The $n$-photon block of its unitary is
\begin{align}
\begin{split}
    &R_{ef}^{(n)}(\theta, \varphi_{ef}) = \begin{pmatrix} 
        1 & 0 & 0 \\ 
        0 & e^{-in(\chi_e + \chi_f) \tau/2} & 0 \\
        0 & 0 & e^{-in(\chi_e + \chi_f) \tau/2}
    \end{pmatrix} \\ 
    &\times\begin{pmatrix}
        1 & 0 & 0 \\
        0 & c_n + i\frac{\delta_{ef}}{\sqrt{\delta_{ef}^2+\epsilon_{ef}^2}}s_n & -ie^{i\varphi_{ef}}\frac{\epsilon_{ef}}{\sqrt{\delta_{ef}^2+\epsilon_{ef}^2}}s_n \\
        0 & -ie^{-i\varphi_{ef}}\frac{\epsilon_{ef}}{\sqrt{\delta_{ef}^2+\epsilon_{ef}^2}}s_n & c_n - i\frac{\delta_{ef}}{\sqrt{\delta_{ef}^2+\epsilon_{ef}^2}}s_n 
    \end{pmatrix},
\end{split}
\label{Q_ef_rotation}
\end{align}
where $\delta_{ef}=n(\chi_f-\chi_e)/2$, $\theta_n = \theta\sqrt{\delta_{ef}^2+\epsilon_{ef}^2}/\epsilon_{ef}$, and $\theta = 2\epsilon_{ef} \tau$. We have ignored corrections from driven Stark shifts due to the transmon drive, which we analyze in Sec.~\ref{transmon_rotation_stark_shift_correction}.

Lastly, $\mathcal{H}_\mathrm{JC}$ generates a Jaynes--Cummings 
rotation unitary whose phase depends on the pulse start time $t_1$. We consider a square pulse turned on for a length 
$\tau$:
\begin{equation}\label{eq:sideband_pulse}
    \epsilon_\mathrm{sb}(t) = 
    \begin{cases}
        \epsilon,&  t_1 \leq t \leq t_1+\tau \\
        0,& \mathrm{otherwise}
    \end{cases}.
\end{equation}
The $\{\ket{en}, \ket{fn}, \ket{g,n+1}\}$ block of the unitary generated during this interval is given by
\begin{widetext}
\begin{equation}
\begin{split}
    &S^{(n)}(\theta, \varphi_\mathrm{sb}, \delta) =
    \begin{pmatrix} 
        e^{-i(\Delta_e+n(\chi_e + \Delta_\mathrm{osc}))\tau} & 0 & 0 \\ 
        0 & e^{-i(\delta_\mathrm{sb}+(n+1)\Delta_\mathrm{osc})\tau} & 0 \\
        0 & 0 & e^{-i(\delta_\mathrm{sb}+(n+1)\Delta_\mathrm{osc})\tau}
    \end{pmatrix} \\
    &\times\begin{pmatrix}
        1 & 0 & 0  \\
        0 & e^{-i\varphi_\mathrm{dp}}(c_n - i\frac{\delta_{\mathrm{sb}}}{\sqrt{\delta_{\mathrm{sb}}^2+(n+1)\epsilon_{\mathrm{sb}}^2}}s_n)  & e^{-i(\varphi_\mathrm{idle}+\varphi_\mathrm{dp})}(-ie^{-i\varphi_\mathrm{sb}}\frac{\sqrt{n+1}\epsilon_{\mathrm{sb}}}{\sqrt{\delta_{\mathrm{sb}}^2+(n+1)\epsilon_{\mathrm{sb}}^2}}s_n) \\
        0 & e^{i\varphi_\mathrm{idle}}(-ie^{i\varphi_\mathrm{sb}}\frac{\sqrt{n+1}\epsilon_{\mathrm{sb}}}{\sqrt{\delta_{\mathrm{sb}}^2+(n+1)\epsilon_{\mathrm{sb}}^2}}s_n)& c_n + i\frac{\delta_{\mathrm{sb}}}{\sqrt{\delta_{\mathrm{sb}}^2+(n+1)\epsilon_{\mathrm{sb}}^2}}s_n
    \end{pmatrix},
\label{final_sb_unitary}
\end{split}
\end{equation}
\end{widetext}
where $\delta_\mathrm{sb} = ((n - d + 1)\chi_f-\delta)/2$, $\theta_n = \theta \sqrt{\delta_{\mathrm{sb}}^2+(n+1)\epsilon_\mathrm{sb}^2}/\sqrt{(\delta/2)^2 + d\epsilon_\mathrm{sb}^2}$, $\theta=2\sqrt{(\delta/2)^2+d\epsilon_\mathrm{sb}^2}\tau$, and
\begin{equation}\label{eq:dynamical_phase}
    \varphi_\mathrm{dp}=(\Delta_f-\Delta_\mathrm{osc}+(d-1)\chi_f+\delta)\tau
\end{equation}
is the dynamical phase. $\varphi_\mathrm{idle}=(\Delta_f-\Delta_\mathrm{osc}+(d-1)\chi_f+\delta)t_1$ is an idle phase. This idle phase depends on the start time $t_1$ of the sideband pulse and arises due to idling the frequency of the drive at the Stark shifted sideband resonance frequency. It can be removed by idling at the undriven resonance frequency and chirping the drive frequency into resonance when the drive is turned on (see Sec.~\ref{sec:chirp}).

The JC control protocol is based on two operations. The first is a rotation between $\ket{g}$ and $\ket{f}$ which we implement by concatenating $\ket{g}$-$\ket{e}$ rotations with $\ket{e}$-$\ket{f}$ rotations:
\begin{equation}
    R(\theta, \varphi) = R_{ef}(\pi, 0)R_{ge}(\theta, \varphi)R_{ef}(\pi, \pi).
\end{equation}
The second operation is a JC rotation 
\begin{equation}
    \mathrm{JC}(\phi, \delta) = S(2\pi, \phi, \delta)
\end{equation}
constrained with $\theta=2\pi$ so that it performs a $2\pi$ rotation on the $\ket{f,d-1}$-$\ket{g,d}$ transition, where $d$ is a chosen cutoff dimension of the oscillator. This construction closes the operations to the Hilbert space of the oscillator up to $d-1$ photons, where the Fock levels at and below can be used to encode a $d$-dimensional qudit. 

\section{System calibration for JC control}\label{sec:system_calib}

In this section, we detail the characterization and calibration procedures for implementing the JC control protocol. All experimental data presented in this paper are obtained using mode three of the weakly coupled multimode bosonic memory in Ref.~\cite{huang2025fast}. The control pulses are generated by a Xilinx ZCU216 FPGA running the Quantum Instrumentation Control Kit (QICK) firmware~\cite{stefanazzi2022qick}. The transmon $\ket{g}$-$\ket{e}$ and $\ket{e}$-$\ket{f}$ drives are produced by one DAC channel, the sideband drive produced by a second channel, and the cavity drive by a third channel, all of which are synchronized to a fixed relative phase at the start of the experiment.

\subsection{System characterization}

The dispersive shifts of the $\ket{e}$ and $\ket{f}$ states are measured through $\ket{g}$-$\ket{e}$ and $\ket{g}$-$\ket{f}$ Ramsey experiments with and without the addition of a photon in the target cavity mode. The photon is prepared by driving the $\ket{f0}$-$\ket{g1}$ transition after preparing the transmon in $\ket{f}$ to implement a SWAP operation.

The driven Stark shift of the sideband interaction is calibrated using spectroscopy. The transmon is prepared in $\ket{f}$ and the sideband drive frequency is swept. The driven $\ket{f0}$-$\ket{g1}$ resonance frequency is shifted from the undriven resonance by the driven Stark shift $\Delta_{f0g1} = \Delta_{f}-\Delta_\mathrm{osc}$.

\begin{figure}
    \centering
    \includegraphics[width=0.85\columnwidth]{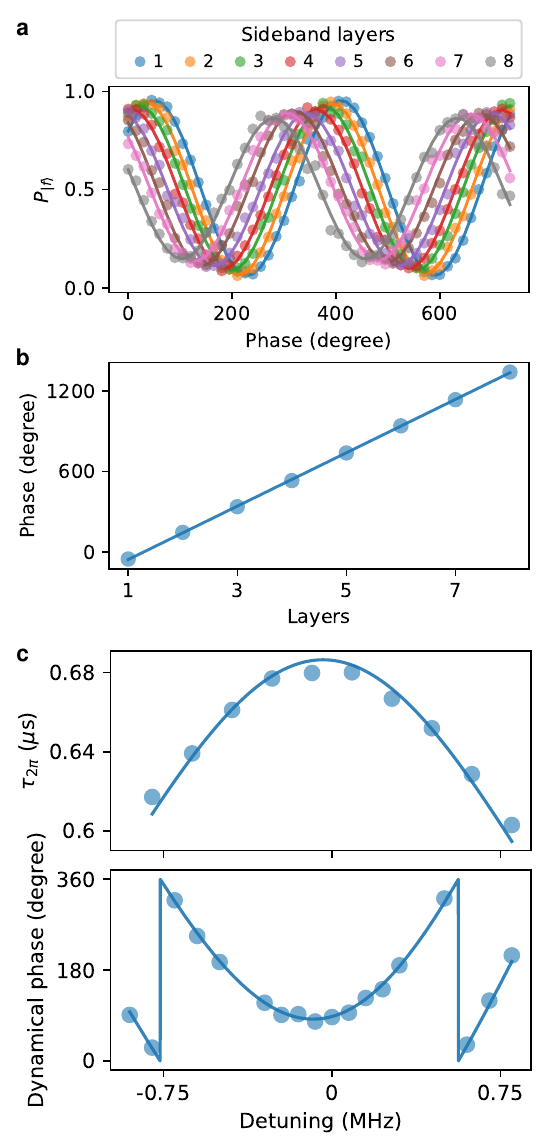}
    \caption{\textbf{Calibration of sideband dynamical phase.} \textbf{(a)} Ramsey experiment with increasing layers of $2\pi$ sideband rotations. The offset phase between sequential layers can be used to extract the dynamical phase. \textbf{(b)} Dynamical phase accumulation as a function of number of $2\pi$ sideband layers for the $\ket{f2}$-$\ket{g3}$ transition. The slope of the fit is the dynamical phase of a $2\pi$ rotation \textbf{(c)} (top) Time of a $2\pi$ rotation as a function of detuning for the $\ket{f0}$-$\ket{g1}$ transition. (Below) The measured dynamical phase as a function of sideband detuning. The solid line is the theoretically predicted dynamical phase from Eq.~\ref{eq:dynamical_phase} based on independently calibrated Hamiltonian parameters.}
    \label{fig:jc_detun_characterization}
\end{figure}

\subsection{Dynamical phase}

For a qudit of dimension $d$, we measure the dynamical phase using Ramsey interferometry between $\ket{g0}$ and $\ket{f,d-1}$. We first prepare $\frac{1}{\sqrt{2}}(\ket{g0}+\ket{e0})$ and then encode the state using a sequence that maps $\ket{g0}\rightarrow\ket{g0}$ and $\ket{e0}\rightarrow e^{-i\varphi}\ket{f,d-1}$~\cite{huang2025fast}, where $\varphi$ is a variable phase swept across experiments. We then apply $N$ layers of $2\pi$ sideband rotations on the $\ket{f,d-1}$-$\ket{g,d}$ transition. Lastly, we decode the state back so that $\ket{f,d-1}\rightarrow\ket{e0}$ and $\ket{g0}\rightarrow\ket{g0}$ and then apply a $\pi/2$ $\ket{g}$-$\ket{e}$ rotation. The Ramsey oscillations for various $N$ are shown in Fig.~\ref{fig:jc_detun_characterization}(a). For each $N$, we find the phase $\varphi_N$ that maximizes the Ramsey fringe. The resulting phase offset between adjacent layers is given by
\begin{equation}
\varphi_{N}-\varphi_{N+1}=\varphi_\mathrm{dp}+d\Delta_\mathrm{osc}\tau+\pi,
\end{equation}
as follows from Eq.~\ref{final_sb_unitary}. Fig.~\ref{fig:jc_detun_characterization}(b) shows $\varphi_N$ 
after subtracting the $\pi$ phase accumulated per layer, as a 
function of the number of resonant ($\delta=0$) $2\pi$ sideband 
layers. We fit the slope to extract the resonant dynamical phase. For sideband rotations with detuning $\delta$, we change the pulse length so that the $\ket{f,d-1}$-$\ket{g,d}$ rotation remains $2\pi$. The lengths of the pulses as a function of detuning are shown in the top plot of Fig.~\ref{fig:jc_detun_characterization}(b). The measured dynamical phase as a function of the sideband detuning $\delta$ is shown in the bottom plot, and agrees well with the analytic expression in Eq.~\ref{eq:dynamical_phase}. The driven oscillator Stark shift $\Delta_\mathrm{osc}$ is small in our device and leads to a small infidelity (see Fig.~\ref{fig:error_budget}(b)), so we ignore this correction. 

Since the gate fidelity is sensitive to the phases applied at each layer, we match the resonant dynamical phase used in the optimizer to the experimentally measured value. To do this, we independently measure the parameters that determine the resonant dynamical phase, $\Delta_f$, $\Delta_\mathrm{osc}$, and $\chi_f$, as shown in Eq.~\ref{eq:dynamical_phase}. We then choose the effective pulse duration $\tau$ used in the optimizer so that the calculated resonant dynamical phase matches the experimental value. For non-square pulses, such as those with ramps, this procedure provides a calibration of the effective pulse duration.

\begin{figure}
    \centering
    \includegraphics[width=0.9\columnwidth]{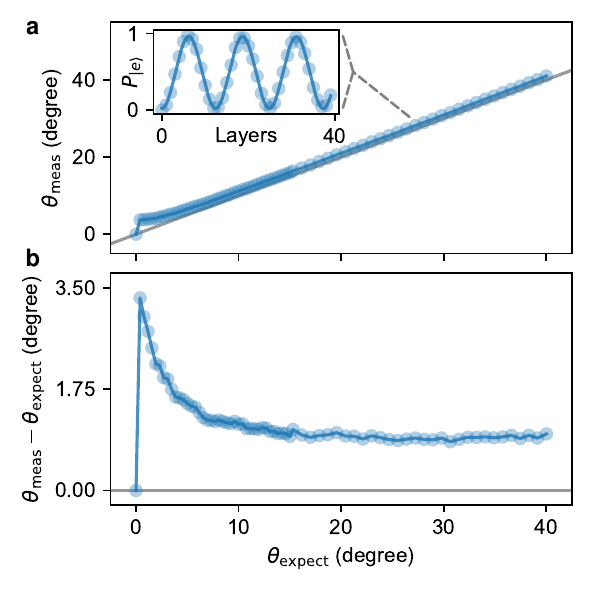}
    \caption{\textbf{Characterization of qubit theta calibration.} \textbf{(a)} Measured versus expected qubit theta rotation angle. The grey line is the ideal calibration. The inset shows the probability of $\ket{e}$ as a function of number of $\theta$ rotations. The period of a sinusoidal fit can be used to extract the value of $\theta$. \textbf{(b)} Difference between the measured and expected theta angle. The error begins to increase below 10 degrees, indicating where the DAC becomes nonlinear. We constrain the optimizer to only optimize over theta angles where the angle is linear.}
    \label{fig:qubit_theta_characterization}
\end{figure}

\subsection{Calibrating ancilla-qubit rotations}

We calibrate arbitrary transmon $\ket{g}$-$\ket{e}$ rotations used for ancilla-qubit rotations in the JC protocol using an error-amplification experiment. We apply a pulse train of target rotations with angle $\theta_\mathrm{expect}$ and measure the probability $P_{\ket{e}}$ of finding the transmon in $\ket{e}$. We fit $P_{\ket{e}}$ as a function of the number of layers in the pulse train (shown in the inset of Fig.~\ref{fig:qubit_theta_characterization}(a)) to a decaying sinusoid to extract the actual rotation angle $\theta_\mathrm{meas}$. This procedure yields a calibration curve for $\theta_\mathrm{meas}$ as a function of $\theta_\mathrm{expect}$. The expected angle $\theta_\mathrm{expect}$ is inferred from the DAC voltage fraction relative to the voltage required for a $\pi$ rotation, which is independently calibrated using a $\pi$-pulse train experiment~\cite{huang2025fast}. The resulting calibration curve is shown in Fig.~\ref{fig:qubit_theta_characterization}(a). We use this curve to convert a target $\theta$ into the appropriate DAC voltage. For sufficiently small $\theta$, the DAC output becomes nonlinear, as shown in Fig.~\ref{fig:qubit_theta_characterization}(b) for angles below $10^\circ$. We therefore constrain the optimizer to use rotation angles larger than this threshold.

\subsection{Calibrating JC rotations}

In our device, JC rotations are implemented through a transmon sideband interaction that couples $\ket{f,n}$ to $\ket{g,n+1}$ for all $n$~\cite{huang2025fast}. We measure the resonance frequency of the $\ket{f,d-1}$-$\ket{g,d}$ transition by performing sideband spectroscopy measurements after preparing the transmon in $\ket{f}$. We start with the cavity in the ground state and locate the $\ket{f0}$-$\ket{g1}$ resonance. Next, we sweep the pulse duration to identify the $\pi$ pulse time corresponding to adding a single photon in the target cavity mode. The process is iteratively repeated for higher photon numbers until we reach the $\ket{f,n}$ to $\ket{g,n+1}$ transition and extract the pulse time required for a $2\pi_{f,d-1\text{-}g,d}$ rotation. 

The FPGA that we use has a fabric clock period of $2.325$ ns, which rounds our pulse times to the nearest integer multiple of this period. This leads to errors in the $2\pi$ rotation and hence process fidelity, as shown in Fig.~\ref{fig:robustness}. We mitigate this error by adjusting the sideband amplitude so that the pulse time for a $2\pi$ rotation is an integer multiple of the clock period. Because we adjust the amplitude, the driven Stark shift also changes and we remeasure the resonance frequency at this new amplitude. 

When the sideband detuning is varied to implement a JC detuning, the sideband Rabi rate is fixed while the pulse length is varied to maintain a $2\pi$ rotation on the $\ket{f,d-1}$-$\ket{g,d}$ transition. For different detunings, the required pulse lengths are generally not integer multiples of the FPGA clock period. This effect can be mitigated by co-adjusting the Rabi rate together with the detuning to bring the pulse length to the nearest integer multiple of the DAC period. While feasible, we do not include this additional calibration overhead in the implementations reported here.

\subsection{Stabilizing the driven Stark shift}\label{sec:stabilizing_stark_shift}

\begin{figure}
    \centering
    \includegraphics[width=\columnwidth]{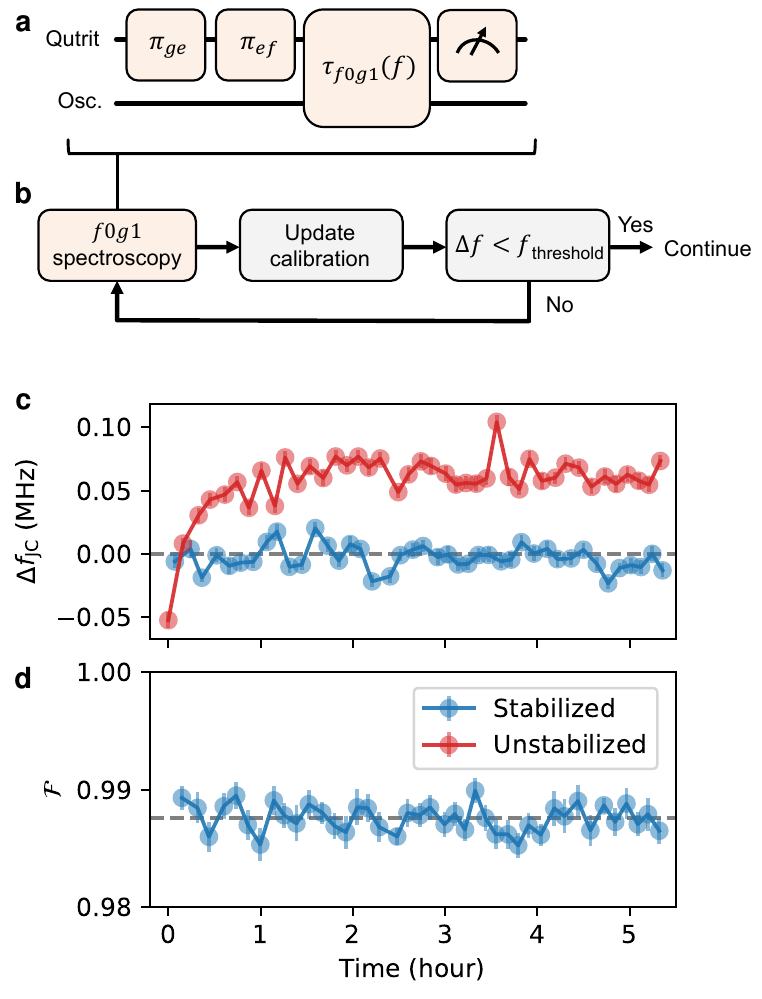}
    \caption{\textbf{Sideband Stark shift stabilization.} \textbf{(a)} Sideband spectroscopy protocol. \textbf{(b)} Protocol for stabilizing the sideband resonance frequency. \textbf{(c)} Measured fluctuations of the $\ket{f}$-state Stark shift with and without stabilization. \textbf{(d)} Fidelity of the $\pi_{f0g1}$ pulse, extracted from the decay of a sideband pulse train with Stark-shift stabilization.}
    \label{fig:sideband stabilization}
\end{figure}

The large Stark shifts induced by the sideband drive increases the gate’s sensitivity to amplitude fluctuations of the drive. We traced these amplitude fluctuations to temperature drifts in the room-temperature electronics. The resulting Stark shift (frequency) fluctuations lead to errors in the JC rotation and phases, such as the dynamical phase. To mitigate this effect, we placed the electronics in a temperature-stabilized rack with $\pm 0.5^{\circ}\mathrm{C}$ stability. However, reducing the frequency fluctuations to $\sim 10~\mathrm{kHz}$ would require temperature stability at the level of $\sim0.05^{\circ}\mathrm{C}$.

We additionally stabilize the Stark shift in software by inserting a calibration measurement prior to each data acquisition. The calibration protocol consists of measuring the Stark shift via spectroscopy of the $\ket{f0}$-$\ket{g1}$ transition and updating the sideband amplitude to maintain a constant Stark shift, as shown in Fig.~\ref{fig:sideband stabilization}(a) and (b). Stabilizing the Stark shift stabilizes the fidelity of the sideband pulse, as shown in Fig.~\ref{fig:sideband stabilization}(d). The fidelity is measured from the decay of a sideband pulse train using the protocol detailed in~\cite{huang2025fast}. 

\section{Deriving the model Hamiltonian}\label{sec:derive_ham}

The device consists of a transmon coupled to a multimode cavity~\cite{huang2025fast}, described by the following multimode Jaynes--Cummings-like system Hamiltonian ($\hbar = 1$):
\begin{align}\label{sys_hamiltonian}
\begin{split}
    \mathcal{H}_{\text{sys}} &=
    \underbrace{\vphantom{\sum_i}\omega_q c^\dagger c - E_J\left( \cos(\varphi) + \frac{\varphi^2}{2}\right)}_{\text{Transmon}} \\
    &\phantom{=}\; + \sum_i \biggl(
    \underbrace{\vphantom{\sum_i}\omega_{\mathrm{osc},i} b_i^\dag b_i}_{\text{Cavity}}
    + \underbrace{\vphantom{\sum_i}g_i ( c - c^\dagger )(b_i-b_i^\dagger)}_{\text{Coupling}}
    \biggr).
\end{split}
\end{align}
All gate operations are activated by driving the transmon. Transmon rotations are implemented by driving near the
$\ket{g}$-$\ket{e}$ and $\ket{e}$-$\ket{f}$ transition frequencies, while sideband operations are implemented by driving near the
$\ket{f,n}$-$\ket{g,n+1}$ transition frequencies of the target cavity mode. We consider the sideband drive given by
\begin{equation}\label{eq:classical_drive}
    \mathcal{H}_\text{d}(t) = - 2i\Omega(t)\cos(\omega_{d} t + \varphi_\mathrm{sb})(c - c^\dagger),
\end{equation}
where $\omega_d$ is the drive frequency, $\Omega(t)$ is a pulse envelope, and $\varphi_\mathrm{sb}$ is a programmable phase. 

In what follows, we make several approximations to derive an effective Hamiltonian that captures the essential physics and enables us to calculate the unitaries for transmon rotations and sideband gate operations. In the experiment, we determine the Hamiltonian parameters of our effective model with experimentally calibrated values.

\subsection{Deriving the Jaynes--Cummings unitary from sideband driving}

We begin by restricting to a single target mode and diagonalizing the transmon-cavity coupling using a dispersive (Schrieffer--Wolff) transformation to first order in $g_i/\Delta_i$, where $\Delta_i=\omega_{\mathrm{osc},i}-\omega_q$ is the detuning between the cavity and the transmon $\ket{g}$-$\ket{e}$ transition. After keeping only up to fourth order in the Josephson nonlinearity, the Hamiltonian is
\begin{equation}
        \mathcal{H} = \omega_q {c}^\dagger {c}
     + \frac{\alpha}{12}\left({c} - \sum_i\frac{g_i}{\Delta_i}{b_i} + \mathrm{H.c.}\right)^4 +\mathcal{H}_\mathrm{d}(t) 
\end{equation}
where $\alpha = -E_c$ is the transmon anharmonicity. We focus on a single cavity mode and drop the mode index $i$.
We then perform a unitary transformation given by $U= e^{\xi(t){c}-\xi^*(t){c}^\dagger}$ to cancel the linear drive, where $\xi(t)=ie^{-i(\omega_dt+\varphi_\mathrm{sb})}\Omega/(\omega_d-\omega_q)-ie^{i(\omega_d t+\varphi_\mathrm{sb})}\Omega/(\omega_d+\omega_q)$.

Finally, we go into the rotating frame of the linearized transmon and oscillator using ${U}_R = e^{-i\left(\omega_q {c}^\dagger {c} + \omega_{c}{a}^\dagger{a} \right)t}$, yielding
\begin{widetext}
\begin{align}
    {\mathcal{H}} &=\frac{\alpha}{12}\left({c} e^{-i\omega_q t} + \frac{g}{\Delta}{a} e^{-i\omega_\mathrm{osc} t} + \eta e^{-i\omega_d t} +  \mathrm{H.c.}\right)^4 \nonumber\\
    &\approx\underbrace{\frac{\alpha}{2}~{c}^{\dagger}{c}\left({c}^{\dagger}{c}-1\right)}_{\text{Transmon anharmonicity}}
+ \underbrace{\frac{K}{2}~ {a}^{\dagger}{a}\left({a}^{\dagger}{a}-1\right)}_{\text{Oscillator self-Kerr}}
+ \underbrace{\vphantom{\frac{a}{b}}\chi_e{c}^\dagger{c} {a}^\dagger{a} }_{\text{Dispersive shift}}
+ \underbrace{\vphantom{\frac{a}{b}}\Delta_q(\eta)~{c}^\dagger{c} + \Delta_\mathrm{osc}(\eta)~{a}^\dagger{a}}_{\text{Driven Stark shifts}} \\
&\quad+
\underbrace{\epsilon_{\rm sb}(\eta)\Big(
e^{-i(2\omega_q-\omega_\mathrm{osc}-\omega_d) t}\,{c}^{2}{a}^\dagger \;+\; e^{i(2\omega_q-\omega_\mathrm{osc}-\omega_d) t}\,{c}^{\dagger 2}{a}
\Big)}_{\text{$\ket{f,n}\leftrightarrow\ket{g,n+1}$ sidebands}}, \nonumber
\end{align}
\end{widetext}
where we have kept only non-rotating terms. Here, $\eta=ie^{-i\varphi_\mathrm{sb}}\Omega\left(\frac{1}{\omega_d-\omega_q}+\frac{1}{\omega_d+\omega_q}\right)$, $\chi_e= 2\alpha\left(\frac{g}{\Delta}\right)^2$ is the cross-Kerr (dispersive) coupling between the oscillator and transmon, and $K=\alpha\left(\frac{g}{\Delta}\right)^4$ is the oscillator self-Kerr interaction.
The sideband rate, as well as the transmon and oscillator driven Stark shifts, depend on the transmon displacement at $\omega_d$ and are given by
$\epsilon_{\rm sb}(\eta)=-\sqrt{2}\alpha\left(\frac{g}{\Delta}\right)\eta$,
$\Delta_q(\eta)=2\alpha|\eta|^2$, and
$\Delta_\mathrm{osc}(\eta)=2\alpha\left(\frac{g}{\Delta}\right)^2|\eta|^2$.
Within these approximations, $\chi_f=2\chi_e$ and $\Delta_f(\eta)=2\Delta_e(\eta)$, which are not quantitatively accurate in experiment due to higher-order corrections. Accordingly, in our modeling and unitary synthesis, we use experimentally calibrated dispersive shifts $\chi_{e,f}$ and driven Stark shifts $\Delta_{e}(\eta)$ and $\Delta_{f}(\eta)$ for the transmon $\ket{e}$ and $\ket{f}$ manifolds.

Truncating the transmon to the lowest three levels ($\{\ket{g},\ket{e}, \ket{f}\}$) and performing an additional unitary transformation ${U}'_R = e^{-i\alpha t~\ket{f}\bra{f}}$ to rotate out the anharmonicity yields the following Hamiltonian in the frame of the undriven transmon and oscillator,
\begin{eqnarray}
    {\mathcal{H}}_{\rm sb} &=& \epsilon_{\rm sb}(t)\left(
    e^{i\varphi_{\rm sb}(t)}\,\ket{g}\bra{f}{a}^\dagger
    \;+\;
    e^{-i\varphi_{\rm sb}(t)}\,\ket{f}\bra{g}{a}
    \right) \nonumber\\
    &&+ \left(\Delta_f + \chi_f{a}^\dagger{a}\right)\ket{f}\bra{f}
    + \left(\Delta_e + \chi_e{a}^\dagger{a}\right)\ket{e}\bra{e} \nonumber\\
    &&+ \Delta_\mathrm{osc}\,{a}^\dagger{a}
    + \frac{K}{2}\,{a}^{\dagger}{a}\left({a}^{\dagger}{a}-1\right),
\end{eqnarray}
where $\varphi_{\rm sb}(t) = \Delta_\mathrm{carrier} t + \varphi_\mathrm{sb}(0)$, $\Delta_\mathrm{carrier} = \omega_d - 2\omega_q - \alpha + \omega_\mathrm{osc}$ denotes the detuning of the applied sideband drive from the undriven
$\ket{f0}$-$\ket{g1}$ transition, and $\varphi_{\rm sb}(0)$ is the starting phase of the sideband tone. We have left the drive-amplitude dependence of the transmon ($\Delta_{e,f}$) and cavity ($\Delta_\mathrm{osc}$) driven Stark shifts implicit, while leaving the sideband drive phase $\varphi_{\rm sb}$ explicit. The drive terms corresponding to transmon $\ket{g}$-$\ket{e}$ and $\ket{e}$-$\ket{f}$ rotations are given in
Eqs.~\ref{ge_drive} and~\ref{ef_drive}.

Ignoring the unused $\ket{e}$ manifold, the sideband Hamiltonian decomposes into independent Jaynes--Cummings doublets $\{\ket{f,n},\ket{g,n+1}\}$. In the subspace of a given doublet, the Hamiltonian reduces to
\begin{align}
    {\mathcal{H}}_{{\rm sb},n}
    &= \bar{\nu}_n\,\mathbb{I}_n
    + \frac{\Delta_n}{2}\,\sigma_{z,n}
    \nonumber \\
    &+ \epsilon_{\rm sb}(t)\left(\sqrt{n+1}\,e^{-i\varphi_{\rm sb}(t)}\,\sigma_{+,n} +\sqrt{n+1}\,e^{i\varphi_{\rm sb}(t)}\,\sigma_{-,n}\right)
\end{align}
where $\bar{\nu}_n =\frac{1}{2}\left(\Delta_f + (2n+1)\Delta_\mathrm{osc} + \chi_f n + K n^2\right)$, $\Delta_n=\Delta_f - \Delta_\mathrm{osc} + (\chi_f - K) n$, $\sigma_{z,n}=\ket{f,n}\bra{f,n}-\ket{g,n+1}\bra{g,n+1}$ and
$\sigma_{+,n}=\ket{f,n}\bra{g,n+1}$. Here $\Delta_{f0g1}=\Delta_f-\Delta_\mathrm{osc}$ is the driven Stark shift of the $\ket{f0}$-$\ket{g1}$ transition.

In our protocol, universal control of a $d$-dimensional qudit is implemented using sideband pulses that are either resonant with the $\ket{f,d-1}$-$\ket{g,d}$ transition or detuned from it by a programmable offset $\delta$. The applied tone must therefore satisfy
\begin{equation}
    \Delta_\mathrm{carrier} = \Delta_{f0g1}(\epsilon_{\rm sb}) + (\chi_f - K)(d-1) + \delta,
    \label{sideband_detuning}
\end{equation}
which follows by transforming to a rotating frame ($R$) that leaves the desired residual detuning $-\delta$ in the $n=d-1$ doublet, while choosing the sideband-drive detuning so that the coupling term is time independent. This transformation is implemented by the unitary ${U}_{R}(t)$, whose form for each doublet $n$ is
\begin{equation}
    {U}^{(n)}_{R}(t)
    = e^{-i\left(\bar{\nu}_{d-1}\mathbb{I}_n + \left(\frac{\Delta_{d-1}+\delta}{2}\right)\sigma_{z,n}\right)t}.
    \label{static_sideband_frame}
\end{equation}

\subsection{Correcting driven Stark shift phases through frequency chirps}\label{sec:chirp}

If the sideband carrier were chosen to maintain the detuning $\Delta_\mathrm{carrier}$ in Eq.~\ref{sideband_detuning} throughout the entire sequence, it would accrue phase in the undriven-Hamiltonian frame in a way that \emph{depends on the absolute timing of the sideband pulse}, including during idle intervals and transmon pulses. To avoid such sequence-position-dependent phases, we choose the sideband carrier to be resonant with the $\ket{f0}$-$\ket{g1}$ transition of the undriven Hamiltonian. To do this, we set $\Delta_\mathrm{carrier}=0$ when the sideband tone is off, so that no phase accumulates when $\epsilon_{\rm sb}=0$. When a sideband pulse is applied in layer $i$, we chirp the tone to the target detuning in Eq.~\eqref{sideband_detuning} by applying a linear phase ramp at the start of the pulse, so that the sideband drive becomes $\epsilon_{\rm sb}(t)\rightarrow \epsilon_{\rm sb}(t)e^{i\varphi_{\rm sb}(t)}$, where
\begin{equation}
    \varphi_{\rm sb}(t) = \Delta_\mathrm{carrier}(t-t_1) + \varphi_{\rm sb}(0),
\end{equation}
and $t_1$ is the pulse start time. This \emph{chirp} procedure introduces a dynamical phase that depends \emph{only on the pulse duration}, not on its start time within the overall sequence, thereby reducing calibration overhead and mitigating coherent phase errors~\cite{naik2017random}.

To obtain the unitary evolution for a sideband-pulse layer following a chirp to the target detuning in the undriven-Hamiltonian frame, consider the evolution given for a pulse turned on at time $t_1$:
\begin{equation}
\begin{aligned}
{U}_{\rm sb}(\tau)
&= \mathcal{T} \exp\!\left[-i\int_{t_1}^{t_1+\tau}{H}_{\rm sb}(t')\,dt'\right] \\
&= {U}_R(\tau)\,e^{-i{H}'_{\rm sb}\tau}\,{U}_R^{\dagger}(0).
\end{aligned}
\label{sideband_unitary}
\end{equation}
Here, $H'_\mathrm{sb}$ is a time-independent Hamiltonian in the frame $U_R$, given by ${\mathcal{H}}'_{\rm sb} = {U}_R^\dagger {H}_{\rm sb}{U}_R - i{U}_R^\dagger \partial_t {U}_R$.

We consider the doublets $\{\ket{f,n}, \ket{g,n+1}\}$ of $\mathcal{H}'_\mathrm{sb}$:
\begin{equation}
\begin{aligned}
{\mathcal{H}}'_{{\rm sb},n}
&=
\bar{\nu}'_n\,\mathbb{I}_n
+ \frac{\Delta'_n}{2}\,\sigma_{z,n}
+ \left[\epsilon_{\rm sb}e^{-i\varphi_{\rm sb}}\,\sigma_{+,n}
+ \mathrm{H.c.}\right] \\
&=
\bar{\nu}'_n\,\mathbb{I}_n
+ \Omega_n\,\vec{\sigma}_n \!\cdot\! \vec{r}_n,
\end{aligned}
\end{equation}
where $\bar{\nu}'_n = \bar{\nu}_n - \bar{\nu}_{d-1}$, $\Delta'_n = \Delta_n - \Delta_{d-1} - \delta = (\chi_f - K)(n-d+1) - \delta$, $\Omega_n = \sqrt{\frac{{\Delta'_n}^2}{4} + (n+1)\epsilon_{\rm sb}^2}$, and $\vec{r}_n = \left( \frac{\epsilon_{\rm sb}\sqrt{n+1}}{\Omega_n}\cos\varphi_{\rm sb}, \frac{\epsilon_{\rm sb}\sqrt{n+1}}{\Omega_n}\sin\varphi_{\rm sb}, \frac{\Delta'_n}{2\Omega_n} \right)$. $\Delta'_n$ is the detuning of doublet $n$, independent of the Stark shift, and reduces to the target detuning $-\delta$ for the $d-1$ doublet. The vacuum state is unaffected and the $\ket{e,n}$ states accumulate phases from the dispersive interaction and driven Stark shifts. The resulting unitary is
\begin{equation}
    \begin{aligned}
        {U}_{\rm sb}(\tau) &= \ket{g,0}\bra{g,0} + \sum_{n=0}^{d-1}e^{-i\left(\Delta_e + n\Delta_\mathrm{osc} + \chi_e n\right)\tau}\ket{e,n}\bra{e,n} \\
        +\sum_{n=0}^{d-1}&\Bigg[\left(\cos(\Omega_n\tau)-i\frac{\Delta'_n\sin(\Omega_n\tau)}{2\Omega_n}\right)e^{-i\varphi_\mathrm{dp}}\ket{f,n}\bra{f,n} \\
&\left(\cos(\Omega_n\tau)+i\frac{\Delta'_n\sin(\Omega_n\tau)}{2\Omega_n}\right)\ket{g,n+1}\bra{g,n+1}\\
        -i&\frac{\epsilon_{\rm sb}\sqrt{n+1}\sin(\Omega_n\tau)}{\Omega_n}e^{-i\left(\varphi_\mathrm{dp}+\varphi_{\rm sb}\right)}\ket{f,n}\bra{g,n+1}\\
        -i&\frac{\epsilon_{\rm sb}\sqrt{n+1}\sin(\Omega_n\tau)}{\Omega_n}e^{i\varphi_{\rm sb}}\ket{g,n+1}\bra{f,n}\Bigg]e^{-i\bar{\omega}_n\tau}.
    \end{aligned}
\end{equation}
Here, $\bar{\omega}_n = (n+1)\Delta_\mathrm{osc}(\epsilon_{\rm sb}) + \big((\chi_f-K)(n-d+1) - \delta\big)/2$ and $\varphi_\mathrm{dp} = \big(\Delta_{f0g1}(\epsilon_{\rm sb}) + \chi_f(d-1) + \delta\big)\tau$. $\bar{\omega}_n$ sets the relative phase accumulated between doublets with different photon numbers and $\varphi_\mathrm{dp}$ is the dynamical phase.
The dynamical phase accumulates due to the detuning of the sideband drive relative to the rotating frame of the undriven system. Chirping the drive to the target frequency ensures that this phase is accumulated solely during the pulse, with its value depending on the Stark and dispersive shifts and the pulse duration $\tau$. The parameter $\varphi_{\rm sb}$ is the programmable phase of the sideband pulse, varied across layers according to the control sequence.

The dominant phase corrections in the unitary above arise from the driven Stark shift of the $\ket{f}$ state, which appears in $\Delta_{f0g1}=\Delta_f-\Delta_\mathrm{osc} \sim 14~\mathrm{MHz}$ and contributes to the dynamical phase, as well as from the dispersive shift $\chi_f \sim 350~\mathrm{kHz}$. The cavity Stark shift, $\Delta_\mathrm{osc} \approx \frac{\chi_e}{2\alpha}\Delta_{f0g1}$, leads to a comparatively smaller frequency correction of $\sim 11~\mathrm{kHz}$. As shown in Figure ~\ref{fig:error_budget}(b), this phase error accounts only for approximately a one percent drop in fidelity for qutrit gates, and we do not include it in the data presented in this paper. In our system, the transmon is weakly coupled to the cavity modes, resulting in a cavity self-Kerr of only $K\sim50~\mathrm{Hz}$, which is therefore neglected.

\subsection{Driven Stark shift corrections for transmon rotations}
\label{transmon_rotation_stark_shift_correction}
We implement transmon $\ket{g}$-$\ket{e}$ and $\ket{e}$-$\ket{f}$ rotations by driving at their respective undriven transition frequencies. Because the transmon is only weakly anharmonic and has nearly oscillator-like matrix elements, these drives also couple neighboring transitions off-resonantly, resulting in driven Stark shifts of the energies. A resonant
$\ket{g}$-$\ket{e}$ drive induces driven Stark shifts of the $\ket{e}$ and $\ket{f}$ levels, while a resonant $\ket{e}$-$\ket{f}$ drive induces driven Stark shifts of the $\ket{g}$, $\ket{e}$, and $\ket{f}$ levels. These driven Stark shifts scale as $\epsilon^2/|\alpha|$ and modify the effective rotation unitaries. We account for these corrections along with the dispersive shift below.

We consider the $\ket{g}$-$\ket{e}$ Hamiltonian in the frame rotating with the drive. Its $n$-photon block $\{\ket{g,n},\ket{e,n},\ket{f,n}\}$ is
\begin{equation}
    \mathcal{H}_{ge}^{(n)} =
    \begin{pmatrix}
        0 & \epsilon_{ge} & 0 \\
        \epsilon_{ge} & n\chi_e & \sqrt{2}\,\epsilon_{ge} \\
        0 & \sqrt{2}\,\epsilon_{ge} & n\chi_e + \alpha_n
    \end{pmatrix},
    \label{eq:H_ge_full}
\end{equation}
where $\alpha_n = \alpha + n(\chi_f-\chi_e)$, $\alpha$ is the transmon anharmonicity, and the factor of $\sqrt{2}$ reflects the scaling of the transmon charge matrix elements. To eliminate the off-resonant coupling to $\ket{f,n}$, we diagonalize the $\{\ket{e,n},\ket{f,n}\}$ block by a rotation of angle $\theta$, where $\tan\theta = 2\sqrt{2}\,\epsilon_{ge}/|\alpha_n|$.

This transformation alters the $\ket{g,n}$-$\ket{e,n}$ coupling to $\epsilon_{ge}\cos\frac{\theta}{2}$ and induces a $\ket{g,n}$-$\ket{f,n}$ coupling of $\epsilon_{ge}\sin\frac{\theta}{2}\approx \sqrt{2}\epsilon_{ge}^2/\alpha_n$. We neglect the latter because it produces energy shifts of order $\epsilon_{ge}^4/\alpha_n^3$. Expanding the dressed energies and matrix elements to second order, and performing an additional unitary transformation ${U} = e^{-i\alpha t \ket{f}\bra{f}}$ to rotate out the bare anharmonicity and move to the undriven frame, we obtain
\begin{equation}
    \mathcal{H}_{ge}^{(n)} =
    \begin{pmatrix}
        0 & \epsilon_{ge} & 0 \\
        \epsilon_{ge} & n\chi_e - \dfrac{2\epsilon_{ge}^2}{\alpha_n} & 0 \\[6pt]
        0 & 0 & n\chi_f + \dfrac{2\epsilon_{ge}^2}{\alpha_n}
    \end{pmatrix}.
    \label{eq:H_ge_eff}
\end{equation}
We then solve for the corrected unitary matrix for Eq.~\ref{Q_ge_rotation}, given by 
\begin{align}
\begin{split}
    &R_{ge}^{(n)}(\theta, \varphi_{ge}) = \begin{pmatrix} 
        e^{-i \delta_{ge}\tau} & 0 & 0 \\ 
        0 & e^{-i \delta_{ge}\tau} & 0 \\
        0 & 0 & e^{-i (n\chi_f + \frac{2\epsilon_{ge}^2}{\alpha_n})\tau}
    \end{pmatrix} \\ 
    &\times\begin{pmatrix}
        c_n + i\frac{\delta_{ge}}{\sqrt{\delta_{ge}^2+\epsilon_{ge}^2}}s_n & -ie^{i\varphi_{ge}}\frac{\epsilon_{ge}}{\sqrt{\delta_{ge}^2+\epsilon_{ge}^2}}s_n & 0 \\
        -ie^{-i\varphi_{ge}}\frac{\epsilon_{ge}}{\sqrt{\delta_{ge}^2+\epsilon_{ge}^2}}s_n & c_n - i\frac{\delta_{ge}}{\sqrt{\delta_{ge}^2+\epsilon_{ge}^2}}s_n & 0 \\
        0 & 0 & 1
    \end{pmatrix},
\end{split}
\label{Q_ge_rotation_stark}
\end{align}
where $\delta_{ge}=n\chi_e/2-\epsilon_{ge}^2/\alpha$.
A similar perturbative analysis can be carried out for the $\ket{e}$-$\ket{f}$ rotation. In this case, correctly capturing the driven Stark shift of $\ket{f}$ requires accounting for the off-resonant driving of $\ket{f}$ to $\ket{h}$ with matrix element $\sqrt{3/2}\,\epsilon_{ef}$. We therefore include the $\ket{h}$ level and incorporate perturbative corrections to second order in $\epsilon_{ef}/|\alpha|$ arising from the off-resonant $\ket{g}$-$\ket{e}$ and $\ket{f}$-$\ket{h}$ drives. The corrected unitary matrix for Eq.~\ref{Q_ef_rotation} is
\begin{align}
\begin{split}
    &R_{ef}^{(n)}(\theta, \varphi_{ef}) \\
    &= \begin{pmatrix} 
        e^{-i\epsilon_{ef}^2\tau/(2\alpha)} & 0 & 0 \\ 
        0 & e^{-i(n(\chi_e + \chi_f)/2 -\epsilon_{ef}^2)\tau} & 0 \\
        0 & 0 & e^{-i(n(\chi_e + \chi_f)/2 -\epsilon_{ef}^2)\tau}
    \end{pmatrix} \\ 
    &\times\begin{pmatrix}
        1 & 0 & 0 \\
        0 & c_n + i\frac{\delta_{ef}}{\sqrt{\delta_{ef}^2+\epsilon_{ef}^2}}s_n & -ie^{i\varphi_{ef}}\frac{\epsilon_{ef}}{\sqrt{\delta_{ef}^2+\epsilon_{ef}^2}}s_n \\
        0 & -ie^{-i\varphi_{ef}}\frac{\epsilon_{ef}}{\sqrt{\delta_{ef}^2+\epsilon_{ef}^2}}s_n & c_n - i\frac{\delta_{ef}}{\sqrt{\delta_{ef}^2+\epsilon_{ef}^2}}s_n 
    \end{pmatrix},
\end{split}
\label{Q_ef_rotation_stark}
\end{align}
where $\delta_{ef}=n(\chi_f-\chi_e)/2-\epsilon_{ef}^2/(2\alpha)$.

We have not included these corrections in the optimizer. They account for approximately 1\% of additional infidelity for the qutrit gates shown in the main text and are significantly mitigated with post-selection (see Fig.~\ref{fig:error_budget}).

\section{Optimizing circuit parameters with gradient descent}\label{sec:optimizer}

\begin{figure*}
    \centering
    \includegraphics[width=0.9\linewidth]{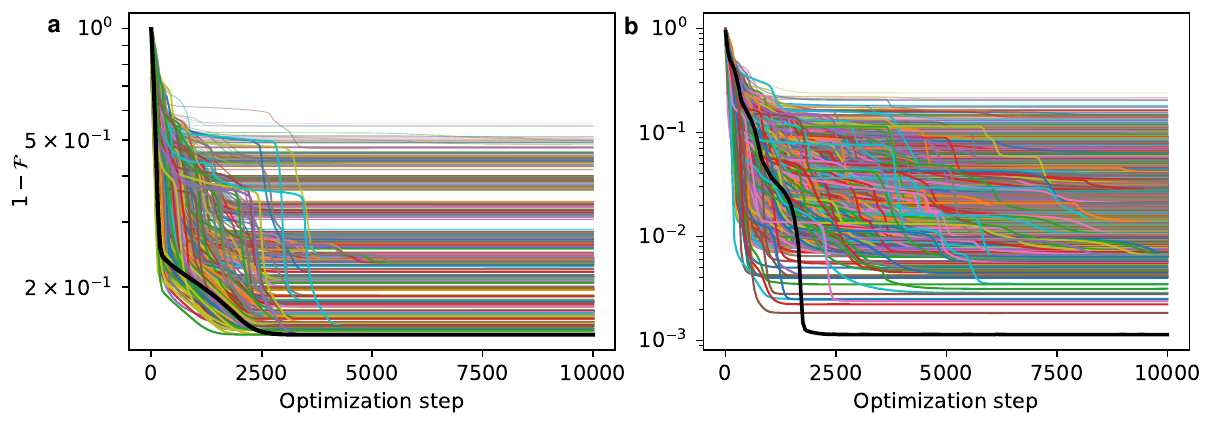}
    \caption{\textbf{Multistart gradient descent convergence.} Fidelity as a function of optimization step for 1000 random initializations, for a qutrit Hadamard gate optimized using \textbf{(a)} 4 circuit layers and \textbf{(b)} 6 layers. For a circuit depth of 4, the fidelity reaches $85.10\%$; a circuit with 6 layers reaches $99.9\%$.}
    \label{fig:multistart}
\end{figure*}

The circuit optimizer is implemented in PyTorch \cite{paszke2019pytorch} and uses the Adam optimizer to perform gradient descent over the circuit parameters (qubit rotation angles and phases, and JC phases and detunings) for a fixed-depth circuit acting on a qutrit-oscillator system. To ensure convergence to local optima, we optimize over a multistart batch of independent initializations simultaneously, as shown in Figure \ref{fig:multistart}. All matrix exponentials, gate compositions, and fidelity evaluations are fully vectorized across the batch dimension, enabling efficient parallel execution on a GPU (NVIDIA GeForce RTX 5070 Ti). The ordered product of the $L$ layer gates is computed using a pairwise tree reduction algorithm, which requires $\mathcal{O}(\log L)$ sequential matrix multiplications rather than $\mathcal{O}(L)$. When optimizing for state transfer, the fidelity is defined as $\mathcal{F}=\left| \bra{\psi_\mathrm{target}}U\ket{\psi_i}\right|^2$, where $\psi_\mathrm{target}$ is the target state, $\ket{\psi_i}$ is the initial state, and $U$ is the optimized unitary. When optimizing a gate, the fidelity is defined as $\mathcal{F} = \left|\mathrm{Tr}(P\,U^\dagger U_\mathrm{target})/\mathrm{Tr}(P) \right|^2$, where $U_\mathrm{target}$ is the target unitary, $U$ is the optimized unitary, and $P$ is a projector onto the subspace of interest. The loss function used is the logarithmic infidelity, $L =  \frac{1}{B}\sum_{j=1}^B\log(1 - F_j)$, averaged across the multistart initializations, where $B$ is the number of independent initializations.

We enforce experimentally motivated constraints during the optimization. In particular, we restrict the qubit rotation angles $\theta$ to a chosen interval to avoid nonidealities associated with the DAC (Fig. \ref{fig:qubit_theta_characterization}). Leakage protection is enforced by construction: each JC layer is closed below the chosen photon-number cutoff so that the dynamics remain confined to the joint subspace spanned by the qutrit states $\{\ket{g},\ket{e},\ket{f}\}$ and the truncated oscillator space $\{\ket{0},\ldots,\ket{d-1}\}$. As a result, all unitaries in the optimizer act on a Hilbert space of dimension $3d$, where $d$ is the qudit dimension.

\section{Robustness to Hamiltonian parameters}
\label{parameter_sensitivity}

\begin{figure}
    \centering
    \includegraphics[width=\columnwidth]{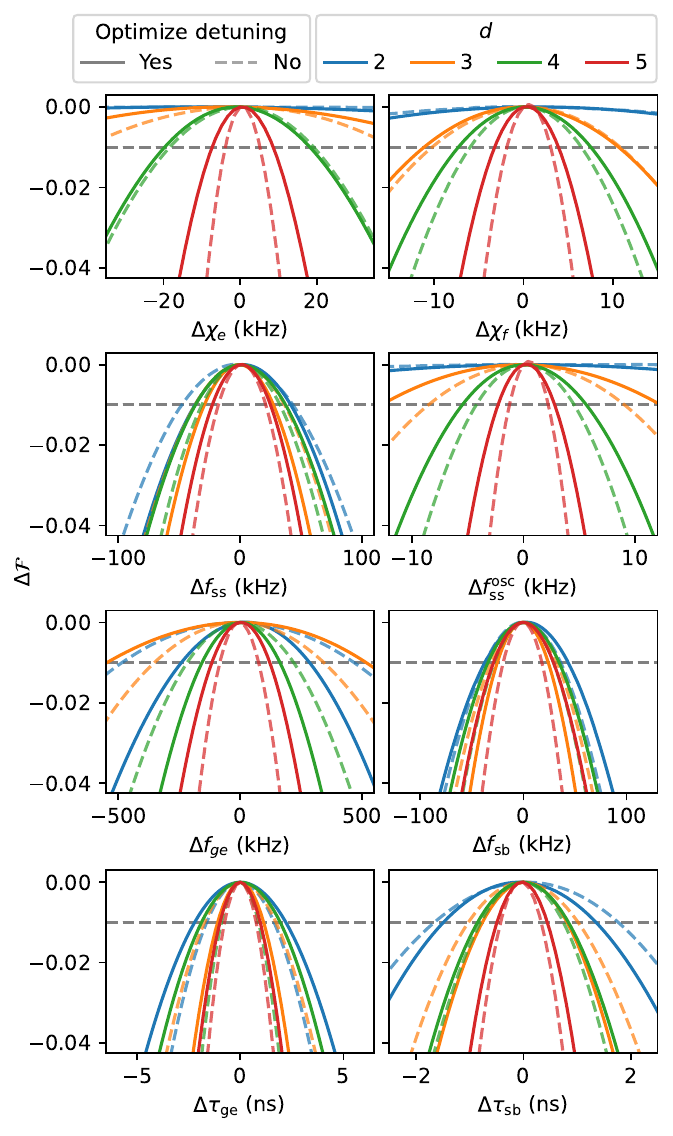}
    \caption{\textbf{Robustness of JC control protocol to Hamiltonian parameters.} Change in process fidelity as a function of Hamiltonian parameter and pulse miscalibration, for a shift gate optimized for $99.9\%$ fidelity. The parameters being swept are the dispersive shifts for $\ket{e}$ ($\Delta \chi_e$) and $\ket{f}$ ($\Delta \chi_f$), the driven Stark shifts for the transmon ($\Delta f_{\rm ss}$) and oscillator ($\Delta f^{\rm osc}_{\rm ss}$), the detunings in the $\ket{g}$-$\ket{e}$ ($\Delta f_{\rm ge}$) and sideband drives ($\Delta f_{\rm sb}$), and the pulse lengths for the $\ket{g}$-$\ket{e}$ ($\Delta \tau_{\rm ge}$) and sideband ($\Delta \tau_{\rm sb}$) drives. The grey line indicates a $1\%$ drop in fidelity.}
    \label{fig:robustness}
\end{figure}

We use numerical simulations of Hamiltonian evolution to determine the sensitivity of gate process fidelities to Hamiltonian parameter and pulse miscalibration. We examine qudit shift gates $X$ optimized for $99.9\%$ fidelity, with and without detuning optimization, using the Hamiltonian in 
Sec.~\ref{app:effective_hamiltonian}. The gate fidelity is computed as detailed in Sec.~\ref{sec:simulated_process_fidelity}, sweeping the dispersive shifts, transmon and oscillator driven Stark shifts, transmon and sideband drive frequencies, and pulse durations. These simulations are performed without decoherence and the results are shown in Fig.~\ref{fig:robustness}. The gate is most sensitive to miscalibrations of the $\ket{f}$ dispersive 
shift $\Delta\chi_f$ and the oscillator Stark shift $\Delta f^\mathrm{osc}_\mathrm{ss}$.

For all Hamiltonian and pulse parameters considered, we find that gate performance becomes increasingly sensitive to miscalibration as the qudit dimension grows. This increased sensitivity arises from both the larger number of circuit layers required to compile higher-dimensional unitaries and the greater susceptibility of larger Hilbert spaces to coherent errors induced by imperfect parameters. Detuning-optimized circuits (solid curves) are consistently more tolerant than fixed-detuning circuits (dashed curves) across all parameters and dimensions, reflecting their lower circuit depth for a given target fidelity.

\section{Qutrit Clifford gates}\label{sec:qutrit_gates}

\begin{figure}
    \centering
    \includegraphics[width=\linewidth]{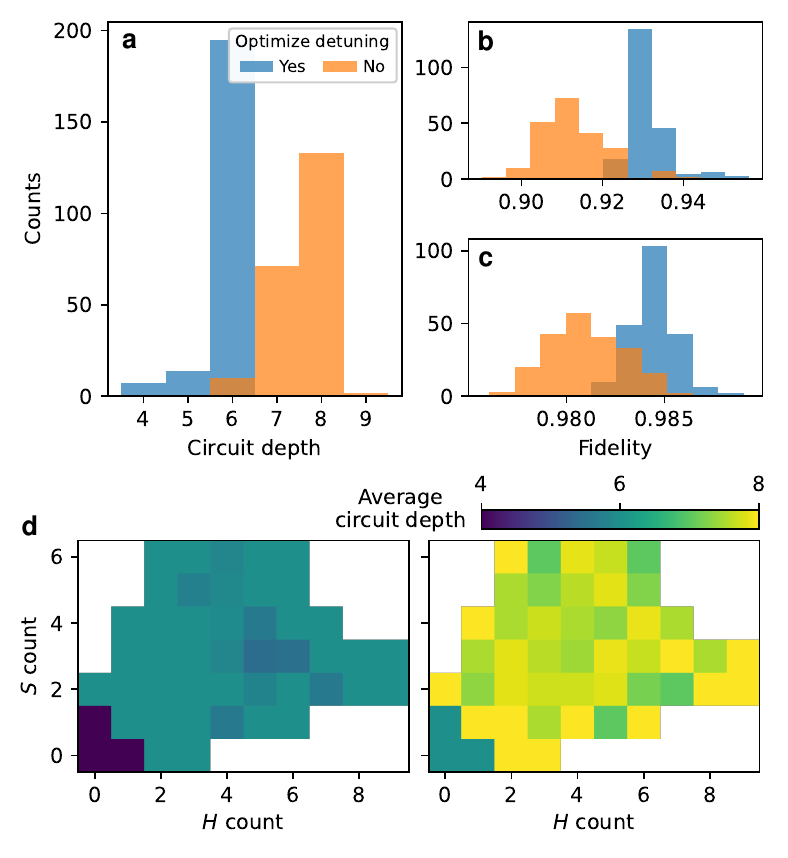}
    \caption{\textbf{Qutrit Clifford gates.} \textbf{(a)} Histograms of the number of circuit layers to implement all 216 qutrit Clifford gates with $99\%$ fidelity, optimized with (blue) and without (orange) JC detuning. Simulated gate fidelities using device parameters, including decoherence, are shown with \textbf{(b)} and without \textbf{(c)} post-selection. \textbf{(d)} Heatmap of the average circuit depth to implement a Clifford gate generated by products of $H$ and $S$, with (left) and without (right) JC detuning optimization. }
    \label{fig:qutrit_cliffords}
\end{figure}

In this section, we study the circuit layers required to directly compile all 216 qutrit Clifford gates using the JC protocol. Fig.~\ref{fig:qutrit_cliffords}(a) shows the circuit layers required to implement all qutrit Clifford gates with $99\%$ fidelity, with and without JC detuning optimization. Fig ~\ref{fig:qutrit_cliffords}(b) and (c) show the process fidelities from Lindblad master equation simulations without and with post-selection respectively. Fig.~\ref{fig:qutrit_cliffords}(d) shows a heatmap of the circuit depths required for qutrit Clifford gates, sorted by the number of $H$ and $S$ gates required to generate it. We find that all the gates can be compiled with similar number of layers. Furthermore, detuning optimization leads to both a lower circuit depth and higher fidelities, when including decoherence.

\section{Ququart and ququint gates}\label{sec:ququart_ququint}

\begin{figure}
    \centering
    \includegraphics[width=\columnwidth]{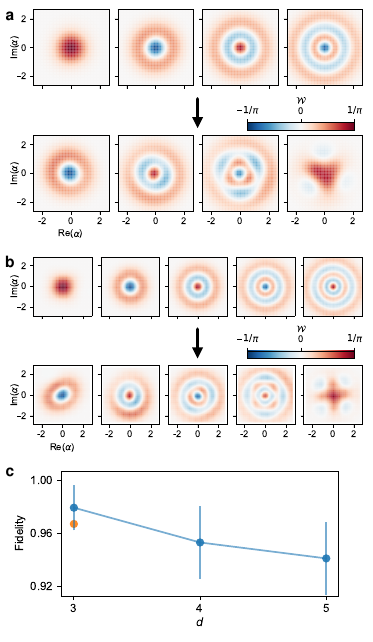}
    \caption{\textbf{Ququart and ququint shift gates.} Experimentally reconstructed Wigner functions for input Fock-basis states (top row) and the corresponding output states after the shift gate (bottom row) for \textbf{(a)} the ququart $X$ gate and \textbf{(b)} the ququint $X$ gate. \textbf{(c)} The blue markers show the post-selected state fidelity averaged over the set of input Fock states. For reference, the orange marker shows the gate fidelity inferred from the qutrit shift gate process matrix.}
    \label{fig:shift_gate_input_output}
\end{figure}

We demonstrate qudit gates beyond a qutrit by implementing a shift gate $X$ for a ququart ($d=4$) and ququint ($d=5$). The experimental Wigner functions of the input and output states for 
the ququart and ququint after the gate are shown in 
Fig.~\ref{fig:shift_gate_input_output}(a) and (b). 

To estimate the gate fidelities for ququart and ququint shift 
gates, we calculate the state fidelity between the measured output state and the output state after acting the ideal shift gate on the measured input state. We use the state fidelity averaged over all input Fock states as a proxy for the gate fidelity. Using this, we obtain gate fidelities of $97.9\%$, $95.3\%$, and $94.1\%$ for the qutrit, ququart, and ququint respectively, shown as blue markers in 
Fig.~\ref{fig:shift_gate_input_output}(c). For the qutrit shift gate, we use the measured process fidelity to infer a gate fidelity using $\mathcal{F}_g = (\mathcal{F}_{\rm proc}d + 
1)/(d+1)$~\cite{horodecki1999generalteleportation}, indicated 
by the orange marker. This agrees well with our state-fidelity-based estimate, validating the gate fidelity proxy for the ququart 
and ququint shift gates.

\section{Comparison of JC control with other protocols}\label{sec:optimal_control_comparison}

In this section we compare the scaling and practical properties of the JC protocol with those of the SNAP-displacement gate set, an alternative approach to universal oscillator control in circuit QED. The SNAP-displacement gate set realizes arbitrary unitaries with a circuit depth that scales as $\mathcal{O}(d)$, for a Hilbert-space dimension $d$~\cite{fosel2020efficient}. This can be understood by noting that each layer contains $d$ independent SNAP phases and two real parameters from the displacement, for a total of $d+2$ real parameters per layer. Since a generic $SU(d)$ unitary is specified by $d^2-1$ real parameters, this suggests that the minimal depth required to implement arbitrary unitaries scales as $\mathcal{O}(d)$.

In contrast, our JC protocol has a circuit depth that scales as $\mathcal{O}(d^2)$. Each layer contributes four tunable parameters (ancilla rotation angle and phase together with sideband phase and detuning), so that realizing $d^2-1$ effective parameters requires a depth of $\mathcal{O}(d^2)$. This is consistent with the scaling observed in our numerics, shown in Fig.~2 of the main text.

While the depth scaling in terms of layer count is larger, the gate time of a sideband layer is much shorter than a typical SNAP layer, with durations that scale as $\sim g/(\epsilon\chi\sqrt{d})$, with the bosonic rate enhancement providing a $\sqrt{d}$ in the denominator. The total sequence time therefore scales as $\mathcal{O}(d^{3/2})$. 

Implementing an arbitrary $d$-parameter SNAP gate efficiently in practice requires optimal-control protocols that can only be 
realized over the shortest allowed timescale on the order of 
$4\pi/\chi$~\cite{landgraf2024fast}. The overall gate time for a 
generic unitary is therefore limited by the dispersive interaction 
rate $\chi$. Comparing typical SNAP gate timescales 
($\sim 2~\mu\mathrm{s}$) to sideband SWAP times 
($\sim 200~\mathrm{ns}$) suggests that the JC protocol has a 
lower gate time until some crossover dimension $d$. The precise 
value depends on the dispersive shift and requires further 
numerical analysis.

The JC protocol also offers practical advantages. By construction, 
it is leakage-protected below the chosen photon cutoff, and all 
primitive pulses come from a discrete, easily calibrated family. 
In contrast, implementing arbitrary SNAP operations in a single 
layer typically requires optimized shaped pulses~\cite{landgraf2024fast} that are more demanding to calibrate and implement. If one instead realizes SNAP gates by compiling separate, spectrally resolved qubit rotations, an arbitrary SNAP would require $d$ serial rotations, driving the layer scaling for the SNAP-displacement protocol back to $\mathcal{O}(d^2)$, making them slower than the JC protocol for a given dimension.

\section{Qubit-oscillator Wigner tomography protocol}\label{sec:qubit_osc_tom}
Joint qubit-oscillator tomography is implemented by performing Wigner tomography on the oscillator state post-selected on the qubit state. In our case, the qubit is defined by the transmon $\ket{g}$ and $\ket{f}$ states. The joint qubit-oscillator density matrix can be written in block form as
\begin{equation}
\rho =
\begin{pmatrix}
S & Q \\
Q^\dagger & P
\end{pmatrix},
\label{eq:block_rho}
\end{equation}
where we have used the tensor product ordering $\mathcal{H}_q\otimes\mathcal{H}_\mathrm{osc}$. Here, $S$, $P$, and $Q$ are operators which act on the oscillator Hilbert space. The diagonal blocks $S=\bra{g}\rho\ket{g}$ and $P=\bra{f}\rho\ket{f}$ describe the oscillator state conditioned on the qubit being in $\ket{g}$ and $\ket{f}$, respectively. The off-diagonal blocks $Q=\bra{g}\rho\ket{f}$ describe the qubit-oscillator coherences. 

To access these blocks experimentally, we first apply qubit rotations $U_j$, then post-select on the qubit being in $\ket{g}$, and then perform Wigner tomography to reconstruct the oscillator state. Different combinations of rotations yield linear combinations of the blocks $\{S,P,Q\}$ in the post-selected oscillator state. The four rotations $U_j\in\{I, \pi_x, \frac{\pi}{2}_x, \frac{\pi}{2}_y\}$ are sufficient to reconstruct the joint qubit-oscillator density matrix.

For a given qubit rotation, the post-selected oscillator state is
\begin{equation}
\rho_{j} = \frac{\widetilde{\rho}_j}{P_j},
\end{equation}
where $\widetilde{\rho}_j=\mathrm{Tr}_q\left(\ket{g}\bra{g}U_j \rho U^\dagger_j\right)$ and the probability of measuring $\ket{g}$ is $P_j=\mathrm{Tr}\left(\widetilde{\rho}_{j}\right)$. Here, the indices $j=1,\ldots,4$ indicate the corresponding qubit rotation. Using our set of rotations, the reconstructed blocks of the joint qubit-oscillator density matrix are then
\begin{align}
S &= P_1\rho_1, \\
P &= P_2\rho_2, \\
Q &= -P_4\rho_4-iP_3\rho_3-\frac{1+i}{2}(S+P).
\end{align}

In the Wigner tomography used to characterize the qutrit gates, we reconstruct the oscillator density matrix including one leakage level above the qutrit (up to and including $\ket{3}$). The tomographic protocol used follows Ref.~\cite{huang2025fast}, where we sample 127 Wigner points with Ramsey contrast correction. For any input state, the leakage during the gate is calculated as the difference in the population outside of the qutrit before and after the gate. The leakage population for all qutrit gates shown in Fig. 4 of the main text is shown in Fig.~\ref{fig:post_selection}(d) and is below $1\%$. For each gate, we have averaged the leakage population over all nine input states used in process tomography. The error bars are calculated by propagating errors in the density matrix reconstruction following Ref.~\cite{huang2025fast} to extract the error in the leakage population.

\section{Process tomography for qutrit gates} \label{app:QPT}

To fully characterize the qutrit gate, we employ quantum process tomography (QPT)~\cite{Zoller1997QPT, Roy2023two_qutrit}. Any linear operation $\mathcal{E}$ can be described with the $\chi$ matrix representation:
\begin{equation}
    \mathcal{E}(\rho) = \sum_{m,n=0}^{d^2-1} \chi_{mn} E_m \rho E_n^\dagger,
\end{equation}
where $d=3$ for a qutrit. We choose the basis $\{E_j\}_{j=0}^{8}$ as the generalized Pauli matrices, providing a set of 9 orthonormal operators shown below.
\begin{align}
        E_0 &= I,  & \nonumber
        E_1 &= X, & 
        E_2 &= Z, \\ 
        E_3 &= X^2, & 
        E_4 &= Z^2, & 
        E_5 &= ZX, \\
        E_6 &= Z^2X, & 
        E_7 &= ZX^2, & 
        E_8 &= Z^2X^2, \nonumber
\end{align}
where 
\begin{equation}
        \mathbb{I} = \begin{pmatrix} 1 & 0 & 0 \\ 0 & 1 & 0 \\ 0 & 0 & 1 \end{pmatrix}, \quad X = \begin{pmatrix} 0 & 0 & 1 \\ 1 & 0 & 0 \\ 0 & 1 & 0 \end{pmatrix}, \quad 
        Z = \begin{pmatrix} 1 & 0 & 0 \\ 0 & \omega & 0 \\ 0 & 0 & \omega^2 \end{pmatrix},
\end{equation}
with $\omega = e^{2\pi i/3}$.
    
The experimental $\chi$ matrix is reconstructed by preparing 9 linearly independent density matrices corresponding to the basis states $\{ \ket{0}, \ket{1}, \ket{2}, \frac{\ket{0}+\ket{1}}{\sqrt{2}}, \frac{\ket{0}+i\ket{1}}{\sqrt{2}}, \frac{\ket{1}+\ket{2}}{\sqrt{2}}, \frac{\ket{1}+i\ket{2}}{\sqrt{2}}, \frac{\ket{0}+\ket{2}}{\sqrt{2}}, \frac{\ket{0}+i\ket{2}}{\sqrt{2}}\}$ and performing state tomography on the outputs obtained after the action of the gate. The gate sequences used to prepare these input states are shown in Table~\ref{table:qudit_process_tomography_input_states}. To account for state-preparation errors, the experimentally measured input states are used in the reconstruction.

We reconstruct the quantum process by finding the Choi matrix $J$ (see Sec.~\ref{app:master_eq_sims} for details) whose associated process $\mathcal{E}_J$ minimizes the sum of squared Frobenius norms between the predicted and measured output states:
\begin{align}
\begin{split}
    \operatorname*{minimize}_{J}&\quad \sum_{k=0}^{d^2-1}\Vert\mathcal{E}_J(\rho^{\mathrm{in}}_k)-\rho^\mathrm{out}_k\rVert_F^2 \\
    \operatorname{subject~to}& \quad J\succeq0, \quad \operatorname{Tr}_2(J)=\frac{I}{d},
\end{split}
\end{align}
where $\mathcal{E}_J(\rho)=d\operatorname{Tr}_1\left((\rho^T\otimes I)J\right)$ and $J\in\mathcal{L}(\mathcal{H}_1\otimes\mathcal{H}_2)$ is the Choi matrix. The constraints on $J$ ensure that the reconstructed process is completely positive and trace preserving.
    
\begin{figure}
    \centering
    \includegraphics[width=\columnwidth]{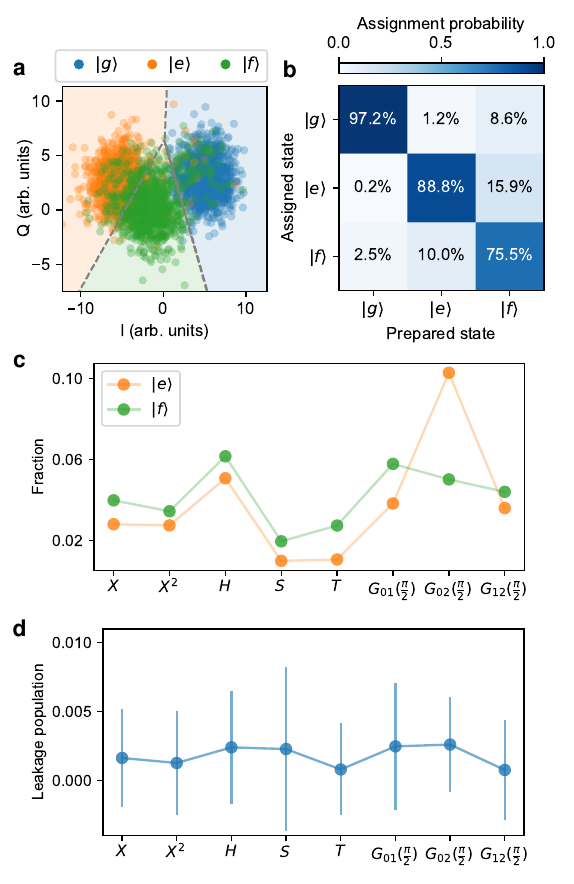}
    \caption{\textbf{Post-selection statistics.} \textbf{(a)} Single shot readout measurements after preparing the transmon in the $\ket{g}$, $\ket{e}$, and $\ket{f}$ states. \textbf{(b)} Three-state transmon readout confusion matrix. \textbf{(c)} Fraction of shots in $\ket{e}$ and $\ket{f}$ discarded by post-selection for all qutrit gates shown in Fig.~4 of the main text. \textbf{(d)} Leakage population (in Fock $\ket{3}$) following each qutrit gate, averaged over all the input states used in process tomography.}
    \label{fig:post_selection}
\end{figure}

The fidelity of the process is then calculated relative to the ideal gate $J_{\rm ideal}$:
\begin{equation}
    \mathcal{F} = \text{Tr}(J_{\rm exp} J_{\rm ideal}).
    \label{eq:process_fidelity}
\end{equation}
The input and output states used to reconstruct the process matrix are oscillator states post-selected on the transmon being in $\ket{g}$. Fig.~\ref{fig:post_selection} shows the fraction of shots that were discarded in the experimental implementation of qutrit gates.  

\begin{table*}
\renewcommand{\arraystretch}{2.25}
\centering 
\begin{tabular}{w{c}{10em} | w{c}{35em}} 
  \hline
  Prepared state & Gate sequence  \\ 
  \hline
    $\displaystyle \ket{n}$  & $\displaystyle\sum_{j=1}^n \left(\pi_{ge}\rightarrow \pi_{ef} \rightarrow \pi_{\ket{f,j-1}-\ket{g,j}}\right)$ \\ 
  \hline
  $\displaystyle\frac{\ket{0}+e^{-i\varphi}\ket{n}}{\sqrt{2}} $ & $\displaystyle \frac{\pi}{2}_{ge}\rightarrow \pi_{ef}(\varphi) \rightarrow \sum_{j=1}^{n-1} \left(\pi_{ge}\rightarrow \pi_{\ket{f,j-1}-\ket{g,j}} \rightarrow \pi_{ge}\rightarrow\pi_{ef}\right)\rightarrow \pi_{\ket{f,n-1}-\ket{g,n}}$ \\ 
  \hline
  $\displaystyle\frac{\ket{m}+e^{-i\varphi}\ket{n}}{\sqrt{2}}$ & JC control protocol \\ 
  \hline
\end{tabular}
\caption{\textbf{State preparation for qudit process tomography.} State preparation sequences used for preparing input states used for qudit process tomography. In the qutrit case, the states prepared using the JC control protocol were implemented in four circuit layers.}
\label{table:qudit_process_tomography_input_states}
\end{table*}

\subsection{Correcting for tomography and offset phases}
\label{app:staticoffsetphase}

The transmon $\ket{g}$-$\ket{e}$, $\ket{e}$-$\ket{f}$, and sideband drives in Eqs.~\ref{ge_drive}--\ref{jc_hamiltonian} contain static offset phases $\varphi_1$, $\varphi_2$, and $\varphi_3$ that arise from the independent phase references of their drive generators. In our device, these drives are generated by separate DACs, each of which initializes with a fixed but arbitrary offset phase. Since these phases can be removed by a unitary transformation, as discussed in Sec.~\ref{app:effective_hamiltonian}, they are not included explicitly in the optimization. The experimentally implemented gate is therefore related to the optimized gate by this unitary transformation. When computing the process fidelity, we account for this freedom by sweeping the corresponding phases of the unitary transformation applied to the ideal reference gate and choosing the phase that maximizes the fidelity.

In the process tomography protocol, we post-select on the transmon being in $\ket{g}$ prior to performing Wigner tomography on the output state following the gate. This measurement imprints a photon-number-dependent phase on the oscillator state and arises from the cross-Kerr interaction between the target oscillator and the readout resonator, which shifts the oscillator frequency when the readout resonator is populated during the post-selection measurement. Calibration errors in the oscillator frequency used for Wigner tomography can additionally contribute photon-number-dependent phase offsets; we mitigate this by performing a cavity Ramsey experiment to calibrate the oscillator frequency to a precision of $\sim 100$~Hz. 

We correct for this phase by applying a unitary transformation $R(\varphi)=\sum_j e^{ij\varphi}\ket{j}\bra{j}$ to the measured output states, where $\varphi$ is an independently measured phase that undoes the cross-Kerr phase accumulated during post-selection. The readout cross-Kerr correction phase is measured by performing Wigner tomography on the prepared state $\frac{1}{\sqrt{3}}(\ket{0}+\ket{1}+\ket{2})$ after $N$ post-selection measurements. We extract the phase accumulated per post-selection measurement from the slope of the measured phase as a function of $N$, which is found to be $-19^\circ$. 

The process fidelities with the post-selection phase correction are shown in the solid blue markers in Fig.~4(g) of the main text, corresponding to a mean post-selected process fidelity of $95.7\%$ across the gate set. If we instead allow $\varphi$ to vary freely and maximize the process fidelity, we extract the hollow blue markers in the figure, corresponding to a mean process fidelity of $96.3\%$. The additional $\varphi$ correction beyond that used for the post-selection phase correction corresponds to an additional phase gate applied following the target gate. This is approximately $-20^\circ$ for the gate set in Fig.~4(g) of the main text. The qutrit gate set with an additional phase gate is still universal. Furthermore, since it is a known phase gate, it can be corrected in subsequent operations. We attribute these residual phases to oscillator Stark shifts from the sideband pulse and Stark shifts from transmon $\ket{g}$-$\ket{e}$ and $\ket{e}$-$\ket{f}$ rotations discussed in Sec.~\ref{sec:derive_ham}, which are not included in the optimizer Hamiltonian used for the data presented here but can be incorporated in future implementations.

\begin{figure*}
    \centering
    \includegraphics[width=\linewidth]{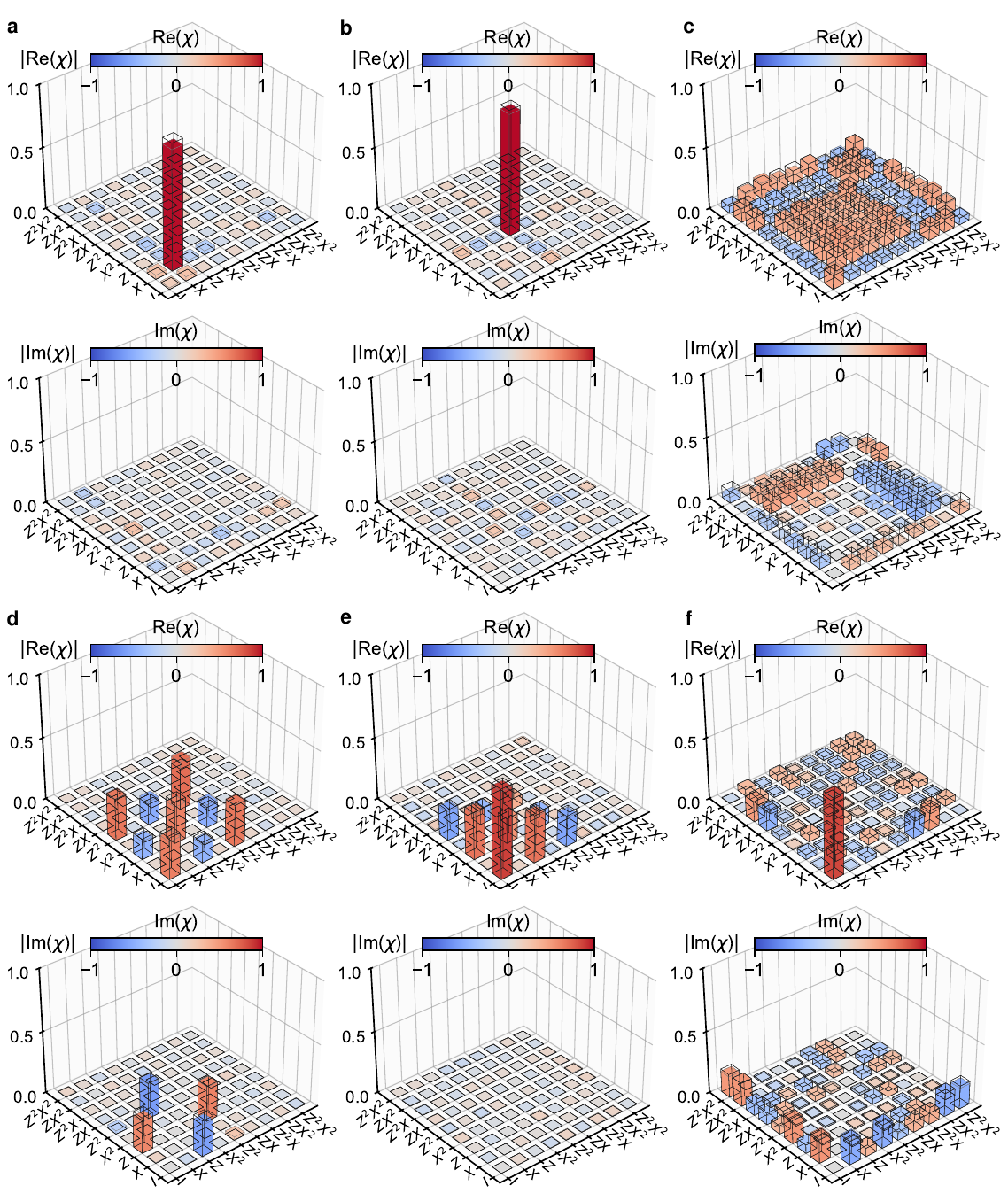}
    \caption{\textbf{Process tomography of universal qutrit gates.} Real and imaginary parts of the process matrices for the qutrit \textbf{(a)} $X$ gate, \textbf{(b)} $X^2$ gate, \textbf{(c)} Hadamard gate, \textbf{(d)} phase $S$ gate, \textbf{(e)} $T$ gate, and \textbf{(f)} $\pi/2$ Givens rotation in the $\ket{0}$-$\ket{1}$ subspace. The solid, black lines indicate the ideal gate.}
    \label{fig:process_matrices_part1}
\end{figure*}

\begin{figure*}
    \centering
    \includegraphics[width=0.69\linewidth]{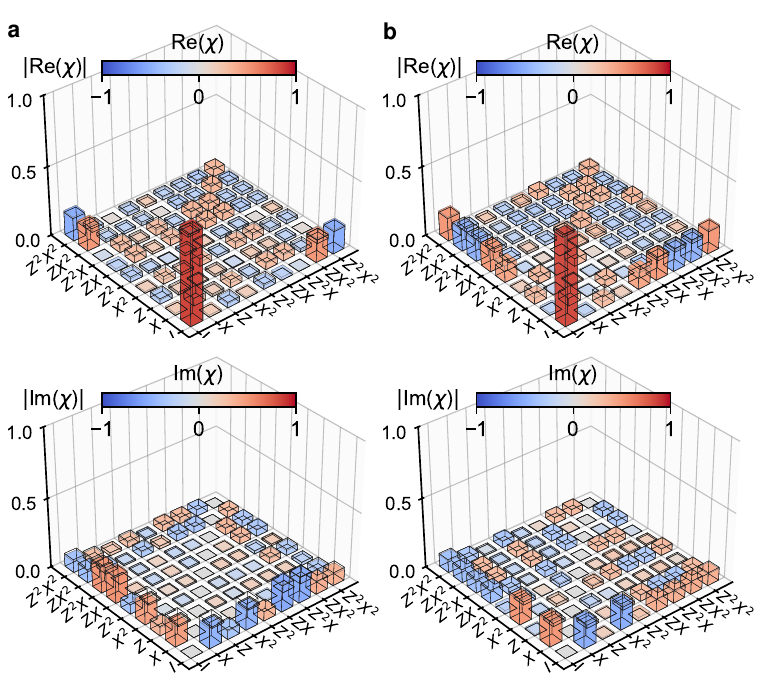}
    \caption{\textbf{Process tomography of universal qutrit gates.} Real and imaginary parts of the process matrices for a \textbf{(a)} $\pi/2$ Givens rotation in the $\ket{0}$-$\ket{2}$ subspace and \textbf{(b)} a $\pi/2$ Givens rotation in the $\ket{1}$-$\ket{2}$. The solid, black lines indicate the ideal gate.}
\label{fig:process_matrices_part2}
\end{figure*}

\section{Lindblad master equation simulations and process fidelity error budget}
\label{app:master_eq_sims}

\begin{figure*}
    \centering
    \includegraphics[width=0.85\linewidth]{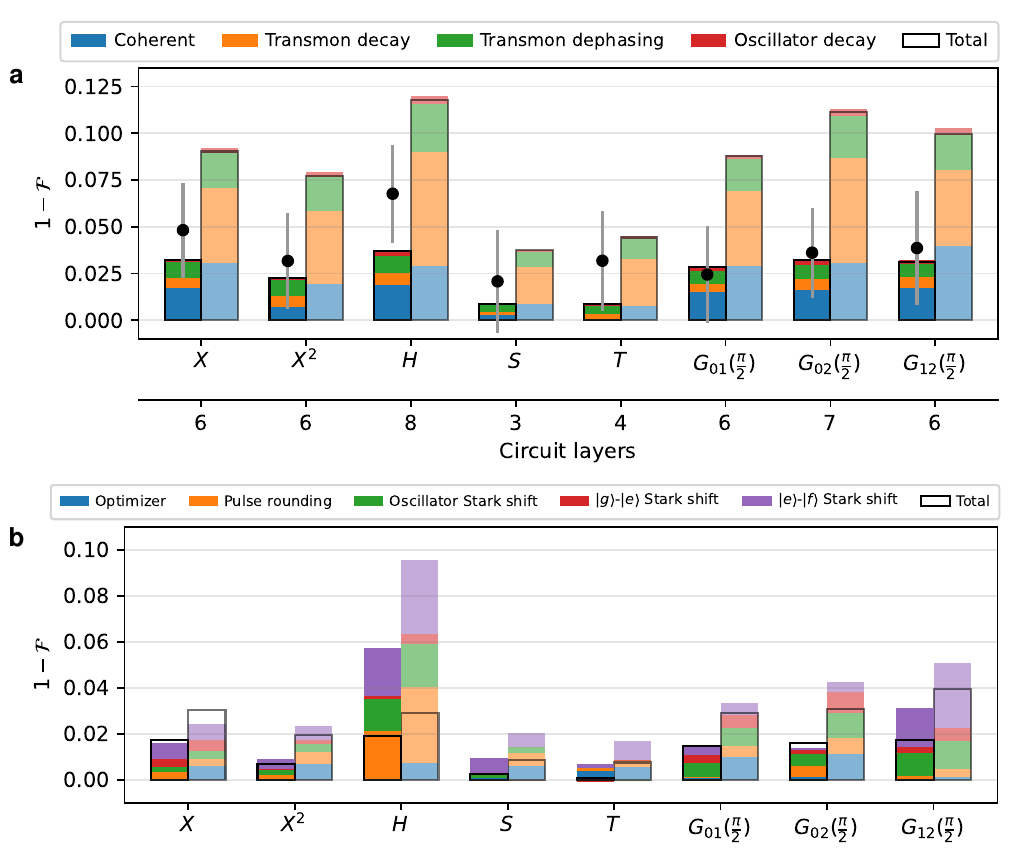}
    \caption{\textbf{Error budget of qutrit gates.} \textbf{(a)} Error budget for qutrit gates from Lindblad master equation simulations. Light colors show the error budget without post-selecting on the transmon being in $\ket{g}$ (right bar), while dark colors show the corresponding error budget with post-selection (left bar). The experimentally measured infidelities are shown in black markers. Without post-selection, the dominant source of infidelity is transmon decay followed by transmon dephasing. Post-selection reduces all infidelities and notably removes a significant fraction of transmon decay errors. After post-selection, transmon dephasing becomes the dominant incoherent source of error. \textbf{(b)} Breakdown of the coherent errors in \textbf{(a)}, including optimizer error, Hamiltonian terms not included in the optimization, and pulse-timing errors from rounding pulse lengths to integer multiples of the fabric clock period of the FPGA used in our experiments. The omitted Hamiltonian terms include $\ket{g}$-$\ket{e}$ and $\ket{e}$-$\ket{f}$ Stark shifts from the transmon drives, and oscillator Stark shifts from the sideband drive.}
    \label{fig:error_budget}
\end{figure*}

We performed Lindblad master equation simulations~\cite{lambert2025qutip5quantumtoolbox} to quantify the effects of decoherence and parameter miscalibration for the JC protocol. The gates are simulated on the pulse level using the Hamiltonian described in Sec.~\ref{app:effective_hamiltonian}. We include decoherence channels corresponding to transmon decay, transmon pure dephasing, and oscillator decay. The jump operators and associated timescales, $\gamma^{-1}$, are taken from Table~S1 of Ref.~\cite{huang2025fast} and summarized in Table~\ref{table:jump_ops}. All experimental data presented in this paper are obtained using mode three of the weakly coupled multimode bosonic memory in Ref.~\cite{huang2025fast}. 

\begin{table}
\renewcommand{\arraystretch}{1.2}
\begin{tabular}{w{c}{13.5em} | w{c}{6.5em}| w{c}{4.5em}} 
  \hline
  Decoherence process & Jump operator & $\gamma^{-1}$ \\ 
  \hline
  Transmon $\ket{e}\rightarrow\ket{g}$ decay & $\ket{g}\bra{e}$ & 55.83 $\mu$s  \\ 
  Transmon $\ket{f}\rightarrow\ket{e}$ decay & $\ket{e}\bra{f}$ & 28.85 $\mu$s  \\ 
  Transmon $\ket{e}$ pure dephasing  & $\ket{e}\bra{e}$ & 81.76 $\mu$s  \\ 
  Transmon $\ket{f}$ pure dephasing  & $\ket{f}\bra{f}$ & 99.94 $\mu$s  \\ 
  Oscillator decay & ${a}$ & 1.298 ms \\
  \hline
\end{tabular}
\caption{\textbf{Lindblad master equation parameters.} Jump operators used in Lindblad master equation simulations and their respective timescales}
\label{table:jump_ops}
\end{table}

\subsection{Process fidelity within a restricted subspace}
\label{sec:simulated_process_fidelity}

Since the optimizer maximizes the unitary fidelity only within a specified computational subspace, we compute the process fidelity on that same subspace while retaining contributions from the larger Hilbert space needed to capture decoherence and leakage errors. We simulate the Lindblad evolution generated by each gate sequence and extract the corresponding transfer matrix. The simulation Hilbert space $\mathcal{H}$ has dimension $n=3(d+1)$, corresponding to three transmon levels and oscillator Fock states up to $d$. The transfer matrix superoperator, which we denote by $\mathcal{E}$, acts on vectorized density matrices of dimension $n^2$ and therefore has dimension $n^2\times n^2$. Because the JC protocol enforces a closed $2\pi$ rotation on the cutoff transition $\ket{f,d-1}$-$\ket{g,d}$, we do not need to keep levels above $\ket{d}$.

The target qudit unitary ${U}_{\rm tar}$ acts on the $d$-dimensional computational subspace, defined by the oscillator Fock states $\{\ket{0},\ldots,\ket{d-1}\}$ with the transmon in $\ket{g}$. During the physical gate sequence, population can transiently occupy higher transmon levels and the boundary oscillator state included in the simulation. We therefore simulate the full evolution in the $n$-dimensional joint transmon-oscillator Hilbert space and compare only the simulated channel restricted to the computational subspace of ${U}_{\rm tar}$. Because the optimizer enforces ${U}_{\rm tar}$ only within this subspace, the action on initial states outside it is unconstrained.

To perform this restriction, we construct an isometry ${V}$ whose columns are the computational basis states embedded in the full Hilbert space $\mathcal{H}$. This isometry maps states from the $d$-dimensional qudit subspace into the full simulation space, while ${V}^\dagger$ projects states in $\mathcal{H}$ back onto the computational subspace. With this isometry, we construct ``embedder'' and ``projector'' superoperators, denoted by $\mathcal{L}_\text{emb}$ and $\mathcal{L}_\text{proj}$, defined as
\begin{align}
    \mathcal{L}_\text{emb} (\rho_\text{sub}) &= \mathrm{vec}\left({V} \rho_\text{sub} {V}^\dagger \right), \\
    \mathcal{L}_\text{proj} (\rho_\text{full}) &= \mathrm{vec}\left({V}^\dagger \rho_\text{full} {V} \right),
\end{align}
where $\rho_{\rm sub}$ and $\rho_{\rm full}$ are density matrices in the qudit subspace and full joint transmon-oscillator Hilbert space, respectively.

To calculate the process fidelity, we first project our simulated transfer matrix to our computational subspace, which is done through the following operation
\begin{align}
    \mathcal{E}_\text{sub} &= \mathcal{L}_\text{proj} \mathcal{E} \mathcal{L}_\text{emb}, \\
    \mathcal{E}_\text{sub} (\rho_\text{sub}) &= {V}^\dag \mathcal{E} {V} \rho_\text{sub} {V}^\dag {V}.
\end{align}
The restricted channel $\mathcal{E}_{\rm sub}$ acts on vectorized density matrices in the $d$-dimensional qudit subspace and therefore has dimension $d^2\times d^2$. We then promote the target unitary ${U}_{\rm tar}$ to the corresponding unitary channel $\mathcal{U}$, which has the same dimension as $\mathcal{E}_{\rm sub}$.

We compute the process fidelity by converting the simulated channel and the target unitary channel to their Choi representations. The Choi representation maps a quantum channel acting on $\mathcal{H}$ to a state in $\mathcal{H} \otimes \mathcal{H}$, referred to as a Choi state, by applying the channel to one half of a maximally entangled state with an auxiliary reference system. For a channel acting on the $d$-dimensional computational subspace, we define
\begin{align}
    J_{\mathcal{E}} &=
    \left(I\otimes\mathcal{E}_{\rm sub}\right)
    \left(\ket{\Phi}\bra{\Phi}\right), \\
    J_{\mathcal{U}} &=
    \left(I\otimes\mathcal{U}\right)
    \left(\ket{\Phi}\bra{\Phi}\right), \\
    \ket{\Phi} &=
    \frac{1}{\sqrt{d}}
    \sum_{j=0}^{d-1}
    \ket{j}\otimes\ket{j},
\end{align}
where $\mathcal{I}$ is the identity channel acting on the auxiliary system and $J$ is the Choi represenation of the quantum map.

\begin{figure*}
    \centering
    \includegraphics[width=\linewidth]{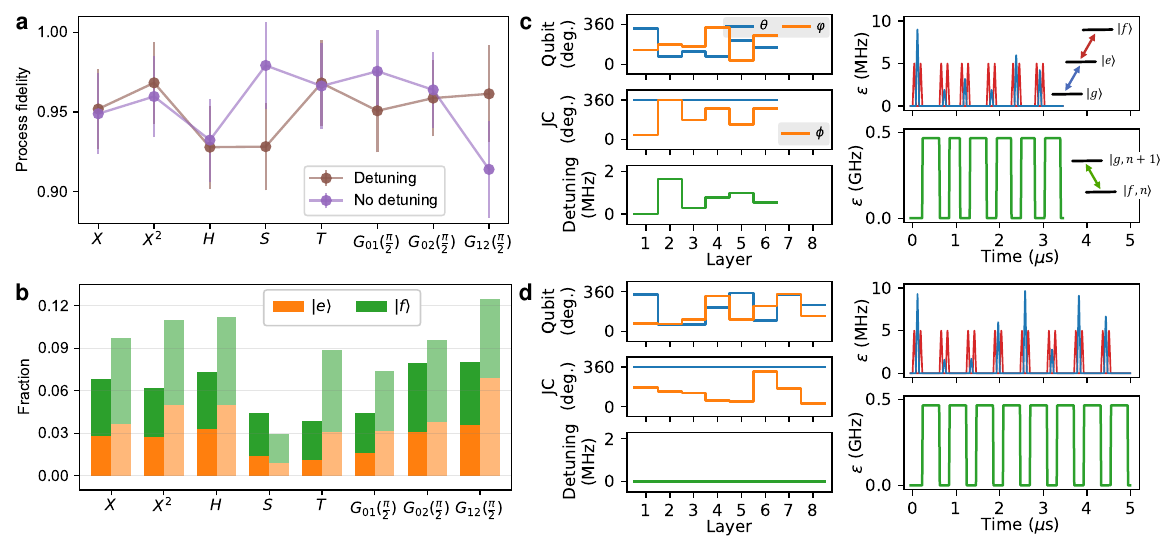}
    \caption{\textbf{Comparison of gates with and without detuning optimization.}\textbf{(a)} Process fidelities for two data sets of universal qutrit gates, implemented with and without Jaynes--Cummings detuning optimization. \textbf{(b)} Fraction of shots in $\ket{e}$ and $\ket{f}$ discarded by post-selection. Light colors show the fractions without detuning optimization, and dark colors show the fractions with detuning optimization. More errors are post-selected out in the longer sequences without detuning optimization. \textbf{(c,d)} (left) Angles of the ancilla qubit and JC rotations, JC detunings, and (right) corresponding pulse sequence for a qutrit shift gate ($X$) optimized for a target fidelity of $99\%$. \textbf{(c)} A 6-layer sequence with detuning optimization and \textbf{(d)} an 8-layer sequence without detuning optimization.}
    \label{fig:detuning_versus_no_detuning_fidelity}
\end{figure*}

For two arbitrary channels, the entanglement fidelity is the Uhlmann fidelity between their normalized Choi states, given by $\mathcal{F}_e =(\operatorname{Tr}\sqrt{\sqrt{\tilde{J}_{\mathcal{E}}}\tilde{J}_{\mathcal{U}} \sqrt{\tilde{J}_{\mathcal{E}}}})^2$. In the case where the target channel $\mathcal{U}$ is unitary, this expression reduces to $\mathcal{F}_e=\operatorname{Tr}\left(\tilde{J}_{\mathcal{E}} \tilde{J}_{\mathcal{U}}\right)$, where $\tilde{J}_{j} = J_{j}/\operatorname{Tr}(J_{j})$. The trace of the Choi matrix $J_{\mathcal{E}}$ equals 1 for a trace-preserving channel and equals the average survival probability  after projection onto the accepted subspace---in our case, the  subspace in which the transmon is found in $\ket{g}$ at the end of the sequence. To model post-selection, we first project the output of the simulated channel onto this accepted subspace before constructing 
$J_{\mathcal{E}}$, and then renormalize by its trace to obtain the conditional post-selected channel $\tilde{J}_{\mathcal{E}}$. The entanglement fidelity computed from this renormalized Choi matrix is the post-selected process fidelity.

The entanglement fidelity is equal to the process fidelity when one of the channels being compared is unitary \cite{nambu2005matrixrepresentationquantumoperations}. The two representations are related by a unitary change of basis.

\subsection{Error budget for qutrit gates}
\label{app:error_budget}

Fig.~\ref{fig:error_budget}(a) shows the error budget compared with the measured process fidelities from process tomography, for the gates shown in Fig. 4 of the main text. The gates that were optimized with detuning are $X$, $X^2$, $T$, and $G_{12}(\pi/2)$; the others were optimized without detuning. The error budget is constructed by running master equation simulations with individual decoherence channels included separately, allowing the infidelity contribution of each channel to be isolated. The agreement between simulation and experiment reflects the advantage of our leakage-protected JC protocol. By confining the dynamics to a known computational subspace, the protocol avoids enhanced photon-loss errors associated with higher Fock levels and reduces sensitivity to uncalibrated higher-order photon-number-dependent nonlinearities. Without post-selection, transmon decay is the 
dominant error source: population in $\ket{f}$ can decay to $\ket{e}$, exiting the computational subspace and remaining dark to subsequent gate dynamics. The lighter and darker bars show the simulated error budgets before and after post-selection, 
respectively. The post-selected bars are substantially smaller across the gate set, demonstrating that heralding on the ancilla returning to $\ket{g}$ removes a significant fraction of relaxation-induced errors. After post-selection, transmon pure dephasing becomes the dominant residual decoherence channel, while coherent errors contribute at the $\sim\!1\%$ level. In the simulations, we do not include cavity decay during the post-selection measurement, 
which we estimate contributes an additional $\sim\!1\%$ to the infidelity. The experimental post-selection statistics are shown in Fig.~\ref{fig:post_selection}.

Fig.~\ref{fig:error_budget}(b) shows the coherent-error budget, which includes residual optimizer error, Stark-shift corrections excluded in the optimization, and pulse-timing errors from rounding pulse durations to integer multiples of the FPGA fabric-clock period. The gates are optimized to a decoherence-free infidelity of approximately $1\%$, leading to a corresponding coherent error. This error contribution is significantly reduced by post-selection. The omitted Hamiltonian terms are the oscillator Stark shift from the sideband drive (Sec.~\ref{sec:chirp}) and the $\ket{g}$-$\ket{e}$ and $\ket{e}$-$\ket{f}$ Stark shifts during transmon rotations (Sec.~\ref{transmon_rotation_stark_shift_correction}). Among these, the largest contributions arise from Stark shifts from the transmon-drive, particularly that associated with the $\ket{e}$-$\ket{f}$ drive used in the composite $\ket{g}$-$\ket{f}$ ancilla rotation. Pulse rounding and the oscillator Stark shift errors vary by gate but generally contribute at or below the percent level.

Fig.~\ref{fig:detuning_versus_no_detuning_fidelity}(a) shows two sets of data corresponding to optimizing with and without Jaynes-Cummings detuning. The best fidelities for each gate are shown in Fig.~4 of the main text. Optimizing with detuning lowers the required circuit layers to achieve a target coherent fidelity, leading to an improvement in the fidelity when including decoherence. However, the qutrit gates implemented are short enough that we do not measure a significant difference in the fidelities between the two sets. As the qudit dimension increases, we expect a more significant improvement in the fidelity when optimizing with detuning (see Fig.~2 of the main text). The advantage of lower circuit depths is reflected in the post selection fraction in $\ket{e}$ and $\ket{f}$, where we measure a lower fraction with detuning optimization, shown in Fig.~\ref{fig:detuning_versus_no_detuning_fidelity}(b). The corresponding pulse sequences for the shift gate with and without detuning optimization are shown in Fig.~\ref{fig:detuning_versus_no_detuning_fidelity}(c) and (d), respectively.

\section{Code availability}

Code for the parameter optimization to implement arbitrary oscillator unitary gates using the JC protocol is available at \url{https://github.com/jordhuang/jc_control}.

\bibliography{references}